\documentclass[a4paper,onecolumn,11pt,accepted=2025-09-04]{quantumarticle}
\pdfoutput=1
\usepackage[utf8]{inputenc}
\usepackage[english]{babel}
\usepackage[T1]{fontenc}
\usepackage{lipsum}
\usepackage{bbm}
\usepackage{amsfonts}
\usepackage{amsmath}
\usepackage{bm}
\usepackage{latexsym}

\usepackage[margin=1.05in]{geometry}
\usepackage{color,graphicx}
\usepackage{array}
\usepackage{enumerate}
\usepackage{amssymb}
\usepackage{amsthm}
\usepackage{pgfplots}
\usepackage{pgf}
\usepackage{tikz}
\usepackage{enumitem}
\usepackage{booktabs}
\usetikzlibrary{patterns}
\usetikzlibrary{arrows.meta}
\definecolor{myblue}{RGB}{0,0,128}

\usepackage{amsthm}
\newtheorem{theorem}{Theorem}[section]
\newtheorem{definition}[theorem]{Definition}

\usepackage{xcolor}
\usepackage{makeidx}
\usepackage[colorlinks=true,linkcolor=blue,anchorcolor=blue,citecolor=red,urlcolor=magenta]{hyperref}
\usepackage{caption}
\usepackage{subcaption}

\usepackage{multirow}
\usepackage{makecell}
\usetikzlibrary{quantikz2}
\usetikzlibrary{arrows.meta}
\def\H{\mathcal{H}}

\newcommand{\tinyspace}{\mspace{1mu}}

\newcommand{\ketbra}[2]{| #1 \rangle\langle #2 |}
\newcommand{\abs}[1]{\left\lVert\tinyspace #1 \tinyspace\right\rvert}
\newcommand{\expval}[1]{\langle #1 \rangle}

\begin{document}

\title{Towards violations of Local Friendliness with quantum computers}

\author{William J. Zeng}
\affiliation{Unitary Foundation}  
\affiliation{Quantonation}  
\email{will@unitary.foundation} 
\homepage{https://willzeng.com/}

\author{Farrokh Labib}
\affiliation{Unitary Foundation}  
\email{farrokh@unitary.foundation} 

\author{Vincent Russo}
\affiliation{Unitary Foundation}  
\email{vincent@unitary.foundation} 
\homepage{https://vprusso.github.io/}

\maketitle 

\begin{abstract}
    Local Friendliness (LF) inequalities follow from seemingly reasonable
    assumptions about reality: (i) ``absoluteness of observed events'' (e.g.,
    every observed event happens for all observers) and (ii) ``local agency''
    (e.g., free choices can be made uncorrelated with other events outside their
    future light cone). Extended Wigner's Friend Scenario (EWFS) thought
    experiments show that textbook quantum mechanics violates these
    inequalities. Thus, experimental evidence of these violations would make
    these two assumptions incompatible. In [Nature Physics 16, 1199 (2020)], the
    authors experimentally implemented an EWFS, using a photonic qubit to play
    the role of each of the ``friends'' and measured violations of LF. One may
    question whether a photonic qubit is a physical system that counts as an
    ``observer'' and thereby question whether the experiment's outcome is
    significant. Intending to measure increasingly meaningful violations, we
    propose using a statistical measure called the ``branch factor'' to quantify
    the ``observerness'' of the system. We then encode the EWFS as a quantum
    circuit such that the components of the circuit that define the friend are
    quantum systems of increasing branch factor. We run this circuit on quantum
    simulators and hardware devices, observing LF violations as the system sizes
    scale. As errors in quantum computers reduce the significance of the
    violations, better quantum computers can produce better violations. Our
    results extend the state of the art in proof-of-concept experimental
    violations from branch factor 0.0 to branch factor 16.0. This is an initial
    result in an experimental program for measuring LF violations at
    increasingly meaningful branch factors using increasingly more powerful
    quantum processors and networks. We introduce this program as a fundamental
    science application for near-term and developing quantum technology.
\end{abstract}

%%%%%%%%%%%%%%%%%%%%%%%%%%%%%%%%%%%%%%%%%%%%%%%%%%%%%%%%%%%%%%%%%%%%%%%%%%%%%%%%%%%%%%%%%%%%%%%%%%%%%%%%%%%%%%%%%%%%%%%
\section{Introduction}
\label{sec:introduction}
%%%%%%%%%%%%%%%%%%%%%%%%%%%%%%%%%%%%%%%%%%%%%%%%%%%%%%%%%%%%%%%%%%%%%%%%%%%%%%%%%%%%%%%%%%%%%%%%%%%%%%%%%%%%%%%%%%%%%%%
Experimental quantum mechanics has long produced evidence that reality differs
from what naive human intuition expects. These experimental results sometimes go
beyond supporting specific quantum mechanical predictions and give evidence
against whole classes of physical theories that obey certain principles. For
example, experimental violations of Bell inequalities provide evidence for
quantum mechanics and show that reality is not described by any theories (even
super-quantum ones) that maintain both local agency and local hidden variables.
These results are part of \emph{experimental
metaphysics}~\cite{shimony1989search, cavalcanti2008reality}, giving evidence
about possible physical theories at the \emph{meta} level.

New tests in experimental metaphysics have been proposed to study the
metaphysical property of \emph{Local Friendliness} (LF)~\cite{bong2020strong}.
Local Friendliness (defined more formally in
Section~\ref{sec:local-friendliness}) is loosely the conjunction of
\emph{objective reality across observers} and \emph{local agency}. Thus, a
violation of the Local Friendliness property provides evidence that one of these
two assumptions need to be jettisoned. Local Friendliness tests are
formalizations and extensions of the Wigner's Friend thought experiment from the
1960's~\cite{wigner1961remarks}. In this work, we:
\begin{enumerate}
    \item propose how quantum computers (and related quantum technology like
    quantum networks) can be used to build increasingly more meaningful (larger
    and loophole-free) tests of Local Friendliness and
    \item use small quantum computers to give experimental evidence (with
    loopholes) of Local Friendliness violations as a first step in this program.
\end{enumerate}

The improvement of quantum technology through academic and industrial
development opens up new avenues for studying fundamental scientific questions.
We are optimistic that a program of Local Friendliness violations can motivate
continued development and benchmarking of quantum technology by testing
important aspects of reality.

This paper proceeds as follows: first in Section~\ref{sec:local-friendliness},
we introduce Local Friendliness inequalities and the Extended Wigner's Friend
Scenario experiments that can be used to violate them. Then, in
Section~\ref{sec:measuring-observers} we propose our experimental program for
increasingly significant LF violations and focus on the branch factor as the
measure of observerness. Next in Section~\ref{sec:experimental-violations}, we
demonstrate violations of LF (with loopholes) at the highest branch factors yet
observed by using quantum computers as an experimental platform. To do this, we
introduce practical approaches to deal with noise in our experiments, including
validating branch factors under noise and reducing the impact of measurement
noise. Then, in Section~\ref{sec:random-unitary-friends} we propose and discuss
approaches for future experiments to obtain higher branch factor violations
with the same number of qubits using different kinds of friend systems. Finally,
we conclude with future directions for advancing the proposed experimental
program.

%%%%%%%%%%%%%%%%%%%%%%%%%%%%%%%%%%%%%%%%%%%%%%%%%%%%%%%
\section{Local Friendliness and tests to violate it}
\label{sec:local-friendliness}
%%%%%%%%%%%%%%%%%%%%%%%%%%%%%%%%%%%%%%%%%%%%%%%%%%%%%%%

Local Friendliness comes from the conjunction of two basic assumptions about
reality~\cite{bong2020strong, haddara2022possibilistic}. We first introduce
these assumptions as metaphysical principles and later define them in a
particular experimental setup for testing Local Friendliness violations.

\begin{definition}[Absoluteness of Observed Events (AOE]
    \label{def:aoe}
    Every observed event is an absolute single event, not relative to anything or anyone.
\end{definition}

AOE states that if an event occurs, it does not occur relative to any particular
observer. This can be viewed as a weaker condition than the intuition violated
by standard relativity. Relativity tells us that two observers may disagree on
the time an event occurs, whereas dropping the assumption of AOE means there
could be a potential disagreement about whether an event occurred. 

\begin{definition}[Local Agency (LA)]
    \label{def:la}
    Free choices are uncorrelated with other events outside their future light cone.
\end{definition}

At a high level, LA can be interpreted to mean that one can construct
independent variables. In other words, LA says that there are events that do not
influence the probabilities of other events. We will need to be able to make
these uncorrelated events in our particular experimental scenarios to show
violations.\footnote{Local Agency can be analyzed into separate assumptions of
Interventionist Causation and a Relativistic Causal Arrow. See
\cite{wiseman2014two} and \cite{wiseman2017causarum} for more details.}

\begin{definition}[Local Friendliness (LF)]
    \label{def:lf}
    The conjunction of AOE and LA.
\end{definition}

It turns out that thought experiments can be designed whereby textbook quantum
mechanics violates the assumption of LF. These thought experiments build upon
the Wigner's Friend thought experiment initially proposed by Wigner
in~\cite{wigner1961remarks} and later refined by Deutsch
in~\cite{deutsch1985quantum}. The implications of these violations being
experimentally confirmed are significant. One is forced to either:
\begin{itemize}
    \item Drop AOE. This means that nature allows separate realities for
    different observers, called \emph{Wigner bubbles} by
    Cavalcanti~\cite{cavalcanti2021view}. There are also other proposals for
    refuting AOE~\cite{ormrod2024quantum, walleghem2024refined,
    vilasini2022general, cavalcanti2023consistency, adlam2022information,
    markiewicz2023relational, rovelli1996relational}. While we are accustomed to
    social and philosophical subjectivity, introducing this subjectivity into
    the heart of physics is a meaningful move.
    
    \item Drop LA. This would be consistent with nonlocal hidden variable
    theories or with some form of superdeterminism~\cite{teufel2009bohmian}.
\end{itemize}
While different interpretations of quantum mechanics support keeping or dropping
either LA or AOE, a meaningful LF violation means one \emph{must} drop one to
remain consistent. Alternatively, one should propose some rule for why one
interpretation should be applied in one scenario and not others. Remaining
generally ``interpretation agnostic" would no longer be tenable.

%%%%%%%%%%%%%%%%%%%%%%%%%%%%%%%%%%%%%%%%%%%%%%%%%%%%%%%%%%%%%%%%%%%%%%%%%%%%%%%%%%%%%%
\subsection{Wigner's Friend and Extended Wigner's Friend Scenario thought experiments} 
\label{sec:wigners-friend}
%%%%%%%%%%%%%%%%%%%%%%%%%%%%%%%%%%%%%%%%%%%%%%%%%%%%%%%%%%%%%%%%%%%%%%%%%%%%%%%%%%%%%%

\begin{figure}[h]
    \centering
    \includegraphics[height=6cm]{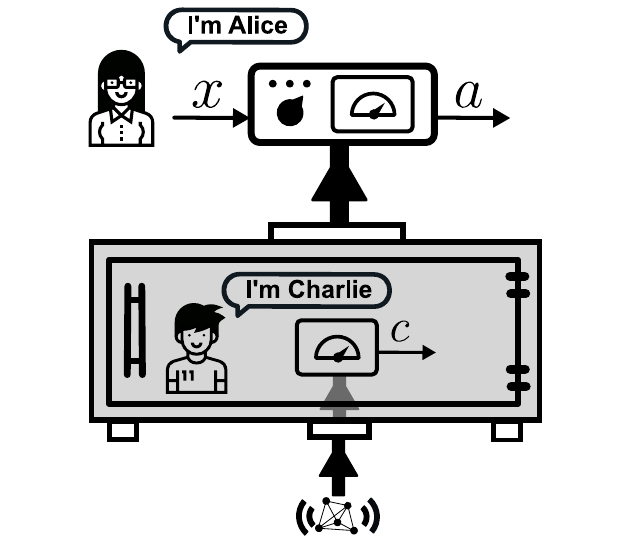}
    \caption{\textbf{Conceptual rendering of the Wigner's Friend scenario.} A system is
    sent into Charlie's sealed lab. Alice has different measurement settings
    labeled by $x$ to observe the sealed lab that contains her friend Charlie
    and his measurement outcome $c$. Alice's measurement outcome has the value
    labeled $a$. Figure replicated and modified with permission from the authors
    from~\cite{bong2020strong}.}
    \label{fig:wf}
\end{figure}

The Wigner's Friend thought experiment (Figure~\ref{fig:wf}) consists of an
observer, Alice, and her friend, Charlie. We assume Charlie is in an isolated
laboratory, receiving and measuring part of an entangled quantum system. Upon
measurement, the outcome obtained by Charlie is known to him but is still not
known to Alice (as she is outside of the isolated laboratory). Charlie then
emerges from his laboratory to inform Alice of the result of Charlie's
measurement. Before Alice receives this information, however, one could
represent Alice's knowledge of the quantum system measured by Charlie as still
in a superposition (even after Charlie has measured and collapsed the quantum
system). It is only after Alice receives the measurement outcome information
from Charlie that her superposition representation of the state prepared and
measured by Charlie collapses to the same measurement outcome that Charlie
previously obtained. If collapse is supposed to be an objective physical
process, then this uncertainty is uncomfortable.

This original thought experiment highlights that it is unclear when the quantum
system collapses (under a Copenhagen interpretation). From Charlie's
perspective, the collapse happened when he applied a measurement to his quantum
system inside the isolated laboratory and obtained his measurement outcome.
Alternatively, from Alice's perspective, the collapse occurred when Alice
obtained information from Charlie about the measurement outcome he obtained.

\begin{figure}[h]
    \centering
    \includegraphics[height=10cm]{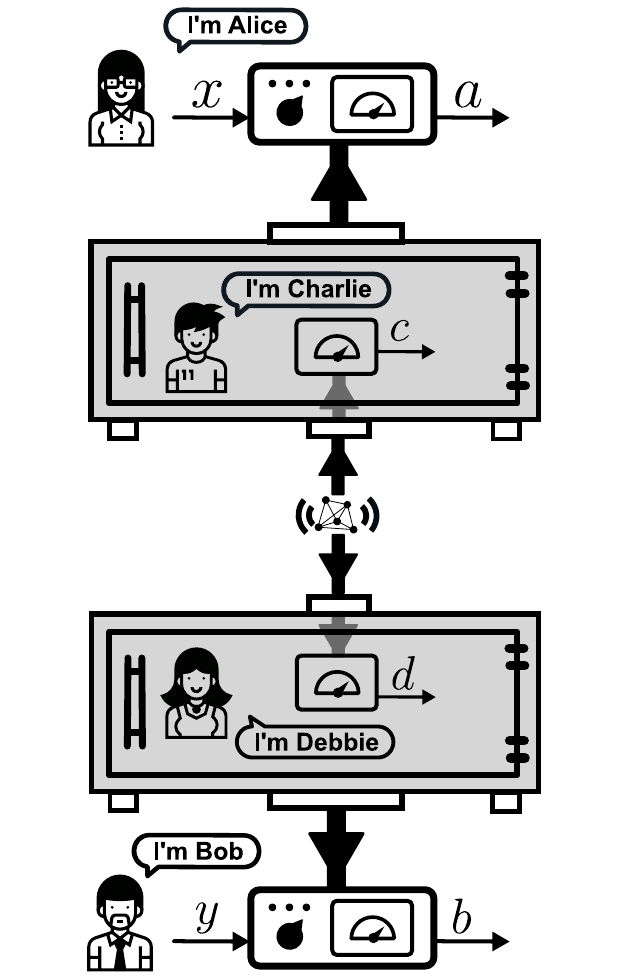}
    \caption{\textbf{Conceptual rendering of an Extended Wigner's Friend Scenario
    (EWFS).} A system is split and sent into two sealed labs. Alice has different
    measurement settings labeled by $x$ to observe the sealed lab that contains
    her friend Charlie and his measurement outcome $c$. Similarly, Bob has
    measurement settings labeled by $y$ for the sealed lab containing Debbie and
    her measurement outcome $d$. Alice's measurement outcome has the value
    labeled $a$, and Bob's has the value labeled $b$. Figure replicated with
    permission from the authors from~\cite{bong2020strong}.}
    \label{fig:ewfs}
\end{figure}

Building on the original Wigner's Friend thought experiment, work by
Brukner~\cite{brukner2017quantum, brukner2018no} and by Frauchiger and
Renner~\cite{frauchiger2018quantum} led to an extension of the thought
experiment called the \emph{Extended Wigner's Friend Scenario} (EWFS),
illustrated in Figure~\ref{fig:ewfs}. One common framing of the extended
Wigner's Friend Scenario comprises parallel instances of the original Wigner's
Friend thought experiment~\footnote{This framing of the EWFS is not the only one
that others have considered. Indeed, there have been others based on the notion
of contextuality~\cite{walleghem2024refined, nurgalieva2023multi,
szangolies2020quantum}. We refer the reader to this review for other such setups
for EWFS~\cite{schmid2023review}.}. We now consider two observers, Alice and
Bob, and their respective friends, Charlie and Debbie. The friends are each
contained in their respective isolated laboratories and possess some quantum
system they share. They each carry out a measurement and subsequently relay
these measurement outcomes to their observer counterparts.

In~\cite{bong2020strong}, the authors performed an experiment with a specific
instance of the EWFS and derived specific LF inequalities that were
experimentally violated. In principle, one could consider generalized extensions
of the EWFS in which there are at least two observers and a corresponding number
of associated friends. Examples of these multiparty scenarios and associated LF
inequalities are studied in~\cite{haddara2024local}. For our purposes, and the
scenario presented in~\cite{frauchiger2018quantum} and~\cite{bong2020strong}, we
restrict our attention to the Extended Wigner's Friend Scenarios of two parties.
We note that, for explanatory purposes, our description makes references to
quantum theory, but the Local Friendliness inequalities are theory-independent.

For the two-party EWFS, the friends Charlie and Debbie share a (not necessarily
entangled) quantum system, where $S_C$ is the part of the system in possession
of Charlie and $S_D$ is the system in possession of Debbie. Charlie and Debbie
are each in isolated labs where they perform a measurement in some specified
basis on their respective part of the system and record the outcome in, say,
their memory. Define $F_C$ as Charlie's laboratory, excluding the system $S_C$
and similarly $F_D$ for Debbie. Charlie's measurement of $S_C$ (in some
specified basis) is, according to the observer Alice, a unitary that acts on
$\H_{F_C} \otimes \H_{S_C}$ where $\H_X$ is the Hilbert space associated to
system $X$, similarly, for Bob and his friend Debbie. 

From Alice and Bob's perspective, there are two ways that they can choose to
measure the quantum system. They can open the lab and simply ``peek'' at the
classical measurement outcome recorded by their friend in the lab.
Alternatively, they can measure the quantum system themselves by reversing the
measurement process that the friends performed, and then directly measuring the
quantum system in a basis of their choosing. They can do this because the
friend's measurement process is unitary, and we assume that Alice and Bob can
manipulate the isolated labs in any way that quantum theory allows.
Interestingly, it can be shown that no-signaling implies that the friends
(Charlie and Debbie) are not only not aware of what outcome they had before
being reversed, but also that they cannot be aware of having been
reversed~\cite{baumann2023observers}.

Unfortunately, performing this experiment with macroscopic observers as friends
is beyond what we could imagine engineering. Instead, we model each friend
(Charlie and Debbie) as a system of qubits. We model measurements performed by
the friends as a CNOT gate. For example, if Charlie measures $S_C$, we apply a
CNOT controlled from $S_C$ to the qubit(s) modeling Charlie. From Alice's
perspective, the measurement is a unitary operation that she can reverse. In the
following section, we illustrate how this produces a violation of Local
Friendliness.

%%%%%%%%%%%%%%%%%%%%%%%%%%%%%%%%%%%%%%%%%%%%%%%%%%%%%%%%%%%%%%%%%%%%%%%%%%%%%%%%%%%%%%%%%%%%%%%%%%%%%%%%%%%%%%%%%%%%%%%
\subsection{Quantum mechanical violations of LF in the three-setting two-party EWFS}
\label{sec:qm-violation-three-setting-two-party}
%%%%%%%%%%%%%%%%%%%%%%%%%%%%%%%%%%%%%%%%%%%%%%%%%%%%%%%%%%%%%%%%%%%%%%%%%%%%%%%%%%%%%%%%%%%%%%%%%%%%%%%%%%%%%%%%%%%%%%%
In this section, we consider the specific details of the two-party and
three-setting EWFS experimentally carried out in~\cite{bong2020strong}. A
conceptual diagram is given in Figure~\ref{fig:ewfs}. Following the EWFS
described in~\cite{bong2020strong}, we consider two observers, Alice and Bob,
who have the option of choosing amongst three possible measurement settings
labeled as $x, y \in \{1,2,3\}$ where $x$ is Alice's setting and $y$ is Bob's
setting. Alice and Bob have respective friends, Charlie and Debbie, who share
some bipartite quantum system and must measure their share of the system in a
basis that depends on the $x$ and $y$ settings, respectively.

If Alice chooses $x = 1$, she peeks and measures Charlie directly (in the
computational basis). If Alice selects $x = 2$ or $x = 3$, she reverses the
measurement operation performed by Charlie and measures the system in a basis
that depends on $x$. The same procedure holds for Bob and his friend Debbie, but
possibly with different basis measurements. To clarify the measurement setting
labels, we will use the convention that $x = 1$ or $y = 1$ is the PEEK setting
and $x \in \{2,3\}$ or $y \in \{2,3\}$ are the REVERSE-1 and REVERSE-2 settings,
respectively.

We can implement the EWFS in a quantum circuit (an example is shown in
Figure~\ref{fig:ewfs-peek-reverse-1}). We are interested in the expectation
values of observables on the subsets of qubits of these quantum circuits that
represent the outcomes for Alice and Bob. We implement the EWFS in a quantum
circuit where we can only do measurements in the computational basis. So,
implementing a measurement in a rotated basis is achieved by simply applying
some rotation gates before measuring in the computational basis. Hence, the
chosen settings $x$ and $y$ determine what circuit to run.

LF inequalities bound what observed outcome statistics are possible for Alice
and Bob when LF holds~\cite{bong2020strong}. For measurement settings $x,y \in
\{1,2,3\}=\{$PEEK, REVERSE-1, REVERSE-2$\}$, we denote $A_x$ and $B_y$ as the
corresponding observables for Alice and Bob. We note that before being derived
in~\cite{bong2020strong}, the LF inequalities were obtained under a different
context regarding device-independent settings~\cite{colbeck2009quantum,
pironio2010random, acin2016certified} in this
thesis~\cite{woodhead2014imperfections} (Appendix-A) under the name of
``partially deterministic polytopes''. Situated between the local and
no-signaling polytopes, these partially deterministic polytopes define the
boundaries of behaviors for which randomness certification is impossible against
a no-signaling adversary, given a specific set of measurements. They serve as a
bridge, connecting the extremes of local and no-signaling constraints in the
landscape of quantum behaviors. This connection was also pointed out
in~\cite{bong2020strong} and has been subsequently investigated
in~\cite{haddara2024local}.

Quantum mechanics indicates that it should, in principle, be possible to violate
these LF inequalities. In~\cite{bong2020strong}, the authors considered a
specific bipartite state and sets of measurements for Alice and Bob that
violated each of the LF inequalities in their work. Our work is exclusively
focused on the so-called \emph{semi-Brukner} inequality which is defined as:

\begin{equation}
    \label{eq:semi-brukner-lf-inequality}
    -\expval{A_1 B_2} + \expval{A_1 B_3} - \expval{A_3 B_2} - \expval{A_3 B_3} - 2 \leq 0.
\end{equation}

We reproduce the specific strategy in~\cite{bong2020strong} that violates these
inequalities. Let $\ket{\psi} \in \H_{S_C} \otimes \H_{S_D}$ be the entangled
quantum state held by Charlie and Debbie is defined as
\begin{equation}\label{eq:ewfs-bipartite-state}
    \frac{1}{\sqrt{2}}\left(|01\rangle - |10\rangle\right) \in \H_{S_C} \otimes \H_{S_D}.
\end{equation}
Additionally, define the projectors $\Pi^{\pm}_{\theta} =
\ketbra{\phi^{\pm}_\theta}{\phi^{\pm}_\theta}$ parameterized by an angle $\theta
\in [0, 2\pi]$ where
\begin{equation}
    \ket{\phi^{\pm}_{\theta}} = \frac{1}{\sqrt{2}}\left(\ket{0} \pm e^{i \theta}\ket{1}\right).
\end{equation}
Define the observable
\begin{equation}\label{eq:ewfs-measurements}
    O_\theta =  \Pi^+_{\theta} - \Pi^-_{\theta}.
\end{equation}

In the middle column of Table~\ref{tab:ewfs-violations}, the upper bound on the
left-hand side of the semi-Brukner inequality is reported using the angles as
described in~\cite{bong2020strong}, while in the right column, the maximum
possible violation is reported by optimizing over the possible angles $\theta$.

\setlength{\tabcolsep}{0.45em}
\begin{table}[htpb]
    \centering
    \begin{tabular}{|c|c|c|}
        \hline
        Inequality & \makecell{LHS of inequality  \\ w/~\cite{bong2020strong} setup} & 
        \makecell{LHS of inequality \\ w/ the optimal angles (Sec.~\ref{sec:experimental-violations})}\\ \hline 
        Semi-Brukner & 0.380364 & 0.82843 \\ \hline            
    \end{tabular}
    \caption{Middle Column: The theoretical results from~\cite{bong2020strong}
    for the left-hand sides of the semi-Brukner inequality. Right Column: The
    theoretical maximum for the left-hand sides of the inequality when we choose
    the optimal strategy as described in Sec.~\ref{sec:experimental-violations}.
    }
    \label{tab:ewfs-violations}    
\end{table}  

%%%%%%%%%%%%%%%%%%%%%%%%%%%%%%%%%%%%%%%%%%%%
\subsection{EWFS scenarios with one friend} 
\label{sec:semi-brukner-no-debbie}
%%%%%%%%%%%%%%%%%%%%%%%%%%%%%%%%%%%%%%%%%%%%
Notice that there is no PEEK setting for Bob in the semi-Brukner inequality, so
he never applies a direct measurement to Debbie, but only measures the system
qubit. In fact, we can show that the semi-Brukner inequality still holds even
when we remove Debbie from the scenario entirely. We prove this statement in
this section. This simplification significantly reduces the resources required
to implement the scenario.

\begin{figure}[!htpb]
    \centering
    \includegraphics[height=8cm]{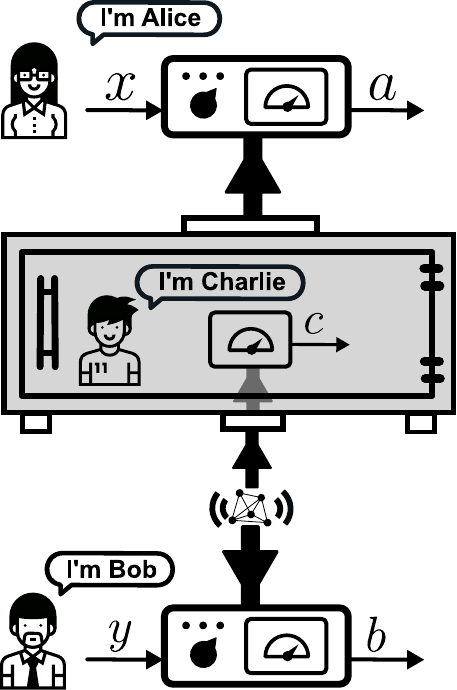}
    \caption{\textbf{Single-friend Extended Wigner’s Friend Scenario.} Alice’s
    friend Charlie, locked in an isolated lab, measures half of an entangled
    pair; Alice may either ``PEEK'' at Charlie’s recorded outcome or undo his
    measurement and re-measure in a rotated basis, while Bob directly measures
    the second half of the pair.}
    \label{fig:ewfs-no-debbie}
\end{figure}

We consider the EWFS scenario where only Alice has a friend (Charlie) while Bob
is on his own. Such a scenario was introduced in~\cite{wiseman2023thoughtful},
where they call it the minimal scenario. The two parties still share an
entangled state on which they apply their respective measurements. Only Alice
has the ability to peek in her lab and see what her friend measured or reverse
the measurement performed by Charlie. This single-friend EWFS variant is
depicted in Figure~\ref{fig:ewfs-no-debbie}.

The assumptions on the probability distribution that give the correlations on
their outcomes should satisfy AOE and LA.~\footnote{Local Agency is the
conjunction of No Superdeterminism and Locality in~\cite{wiseman2014two}.}. Let
$\phi(ab|xy)$ be the empirical distribution of Alice and Bob's outcomes $a,b$
given their measurement choices $x,y$. AOE and LA apply constraints on the
distribution $P(abcd|xy)$ given the measurement choices $x,y$.

\begin{center}
    \label{tab:assumptions}
    \begin{tabular}{@{}p{0.08\textwidth}p{0.85\textwidth}@{}}
    \toprule
    \textbf{AOE:} & $\phi(ab|xy) = \sum_{c}P(abc|xy)$ for all $a,b,x,y$ \\
                  & $P(a|c, x=1, y) = \delta_{a,c}$ for all $a,c,y$ \\[0.5em]
    \textbf{LA:} & $P(c|xy) = P(c)$ for all $c,x,y$ \\
                 & $P(a|cxy) = P(a|cx)$ for all $a,c,x,y$ \\
                 & $P(b|cxy) = P(b|cy)$ for all $b,c,x,y$ \\
    \bottomrule
    \end{tabular}
\end{center}

These are the same requirements on $P$ as in~\cite{bong2020strong}, but without
Bob's friend Debbie. Following the same derivation there, the LF correlations
where only Alice has a laboratory with a friend, are characterized by
\begin{equation}
    \phi(ab|xy) = 
    \begin{cases}
        \sum_{c} \delta_{a,c}P(b|cy)P(c) & \text{if } x = 1 \\
        \sum_{c} P_{\text{NS}}(ab|cxy)P(c) & \text{if } x \neq 1
    \end{cases}.
\end{equation}
Using these correlations, we can give a proof of the semi-Brukner inequality.
The detailed proof can be found in
Appendix~\ref{sec:semi-brukner-no-debbie-derivation}. In our experiments, we
still have a register for Debbie, but we restrict that to only having size one.
All statements in this section still hold in that case too.

%%%%%%%%%%%%%%%%%%%%%%%%%%%%%%%%%%%%%%%%%%%%%%%%%%%%%%
\section{Experimental Program for Local Friendliness Violations with Significant Friend-Observers}
\label{sec:measuring-observers}
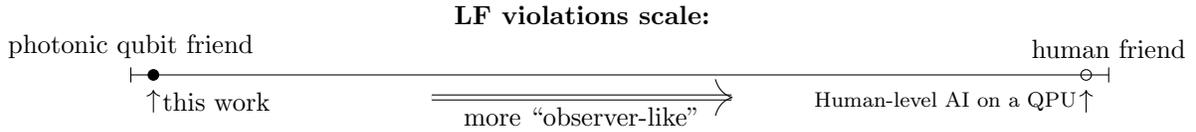
\begin{figure}[!ht]
    \centering
    \begin{tikzpicture}
      \draw (0,0) -- (13,0);  
      \draw (0,-0.1) -- (0,0.1) node[above] {photonic qubit friend};
      \draw (13,-0.1) -- (13,0.1) node[above] {human friend};
      \draw [double, double distance=1.5pt,->] 
            (4,-0.3) -- (8,-0.3) node[midway, below] {more ``observer-like''};
      \node [align=center] at (6,0.8) {\textbf{LF violations scale:}};  
      \draw [->] (0.3, -0.5) -- (0.3, -0.2) node[midway, right] {this work};
      \filldraw [black] (0.3, 0) circle (2pt);
      \draw [->] (12.7, -0.5) -- (12.7, -0.2) node[midway, left] {\footnotesize Human-level AI on a QPU};
      \draw (12.7, 0) circle (2pt);
    \end{tikzpicture}
    \caption{A program designed to test Local Friendliness violations on a
    progressively larger scale increases the significance of the observers used
    as friends. Our work gives the first experimental violations of Local
    Friendliness with friends larger than a photonic qubit.
    In~\cite{wiseman2023thoughtful}, the authors estimate the size of a QPU
    needed to perform experiments using a human-level artificial intelligence
    simulated on a QPU as the friend observer at approximately $10^{19}$ logical
    qubits and logical depth of $10^{14}$ operations. We mark this as a white
    dot as it is a theoretical target while our work gives experimental results
    (black dot).}
    \label{fig:gradient}
\end{figure}

%%%%%%%%%%%%%%%%%%%%%%%%%%%%%%%%%%%%%%%%%%%%%%%%%%%%%%
The inequality violations described in Section~\ref{sec:local-friendliness} are
only significant if the friends in the EWFS count as observers. In a sense,
their status as observers makes the measurement outcome they observe real. While
~\cite{bong2020strong} showed violations with a single photonic qubit as each
friend, one may reasonably doubt that photonic qubits are observers and,
therefore, doubt that the experiment is significant. While running these
experiments with a human as the observing friend would produce the least
controversial outcome, we don't know how to do this with current technology.
Instead, one can design an experimental program to build up to this scale
(Figure~\ref{fig:gradient}).

%%%%%%%%%%%%%%%%%%%%%%%%%%%%%%%%%%%%%%%%%%%%%%%%%%%%%%
\subsection{Quantifying Observerness}
\label{sec:other-observer-measures}
%%%%%%%%%%%%%%%%%%%%%%%%%%%%%%%%%%%%%%%%%%%%%%%%%%%%%%
To establish an experimental program for more ``observer-like'' friends one
needs to choose a physical property of a friend that ``gets closer in scale to a
system that is unambiguously an observer.” Since there is no clear consensus on
a single dimension that defines an observer, there are several dimensions that
could be considered:
\begin{itemize}
    \item More mass~\cite{fein2019quantum, delic2020cooling}.
    \item More complexity~\cite{grinbaum2010quantum, muller2020law}.
    \item More objectivity~\cite{chisholm2021witnessing} (e.g. redundancy and consensus via Quantum Darwinism~\cite{zurek2009quantum}).
    \item Higher degree of irreversibility~\cite{manikandan2019fluctuation, jayaseelan2021quantum, schwarzhans2023quantum}.
    \item More conscious~\cite{wigner1961remarks, shimony1963role, stapp1999attention, hameroff2014consciousness, neven2024testing, bayne2024tests}.
    \item More agency~\cite{dunjko2016quantum, neven2021robots}.
    \item More life, according to a definition like the assembly index~\cite{sharma2023assembly}.
    \item More thoughtful (using a quantum computer that runs a reversible simulation of a human mind)~\cite{wiseman2023thoughtful}.
    \item \textbf{Higher branch factor~\cite{taylor2023wavefunction} (the focus of this work: Section~\ref{sec:branchiness})}.
\end{itemize}

Fundamentally, these dimensions emerge because today there is only one kind of
intuitively unambiguous observer: a human that is heavy, complex, is objectively
described, whose operation appears highly irreversible, is conscious, is alive,
has thoughts, and possesses agency. Observers are special and not equivalent to
other quantum systems~\cite{brukner2021qubits}. In short, different perceptual
states are always macroscopically distinguishable and likely to have a high
branch factor. This means that to violate LF, one needs to scale EWFSs to
include observers beyond toy systems. Still, it would be surprising if one must
get to an actual human to possess an observer. As John Bell inquired, ``Was the
wave function waiting to jump for thousands of millions of years until a
single-celled living creature appeared? Or did it have to wait a little longer
for some highly qualified measurer—with a PhD?"~\cite{bell2001quantum}.

Of the options, this work focuses on the branch factor as it is a quantifiable
measure that can be applied to different systems that can be prepared readily on
available quantum computers. We also note previous
work~\cite{chisholm2021witnessing} that used quantum computers to simulate the
emergence of objectivity as indicative of another direction that is compatible
with quantum computers as an experimental platform.

Mass is also a compelling example of the other proposed dimensions, with several
research groups working on establishing progressively larger superpositions.
These are worth investigating but will require custom experimental setups, e.g.,
the proposed Macroscopic Quantum Resonators MAQRO experiments that aim to test
gravitational decoherence~\cite{kaltenbaek2022maqro}. While the branch factor
has the advantage that it can be run on today's quantum computers, using a
universal quantum computer is overkill, as one needs only to run one program:
unitary measurement and its reverse. In the future, specifically designed EWFS
setups can likely be designed to maximize the branch factor.

%%%%%%%%%%%%%%%%%%%%%%%%%%%%%%%%%%%%%%%%%%%%%%%%%%%%%%
\subsection{Branch factor as a measure of observerness}
\label{sec:branchiness}
%%%%%%%%%%%%%%%%%%%%%%%%%%%%%%%%%%%%%%%%%%%%%%%%%%%%%%

In the EWFS, a quantum system in (up to normalization) a superposition of states
$\ket{\psi}_0 + \ket{\psi}_1$ interacts with a friend in state
$\ket{F}_{\textit{init}}$ resulting in the combined state $\ket{\psi}_0\ket{F}_0
+ \ket{\psi}_1\ket{F}_1$. The pointer states $\ket{F}_0$ and $\ket{F}_1$ are the
states of the friend after having measured the system state. For our
experimental program, we need a measure on these pointer states that describes
how well they are acting as classical observer states. We propose using the
\emph{branch factor}, as described in~\cite{taylor2023wavefunction}, to quantify
this ``observerness''. 

We argue that branch factor is a natural information-theoretic metric to
quantify observerness. As we review in the definitions below, a high branch
factor means that operationally determining whether the friend is in a
superposition of the two states or in a classical mixture of the two states
would have high complexity. The hardness of distinguishing the superposition
from a classical mixture should be sufficient for labeling a system as
classical.

The idea is that pointer states (also called branch states) are classical when
they (i) are easy to distinguish by measurements, and (ii) are hard to
interfere. Interfering the branch states would operationally distinguish between
there having been a classical mixture of states instead of an underlying
superposition. Thus, we want interference to have high complexity. As described
in~\cite{taylor2023wavefunction}, these conditions are sufficient to ``allow the
full state to be replaced with a probability distribution over the branches (a
mixed state) for the purposes of natural time evolution and local measurements,
so that results can be calculated separately for each branch and added
probabilistically.

To mathematically define the branch factor, we first consider two complexity
metrics on interference and distinguishability as defined
in~\cite{aaronson2020hardness}. Let $C(U)$ be the circuit complexity of some
unitary, i.e., the minimum number of one- and two-qubit unitaries required to
perform $U$. Here, single-qubit unitaries have weight one and two-qubit
unitaries have weight two. Let $\ket{\psi_0}$ and $\ket{\psi_1}$ be two
orthogonal quantum states.

\begin{definition}[Interference complexity proxy]
    \label{def:interference-complexity}
    Let $0 \leq \delta \leq 1$. The interference complexity $C_I(\ket{\psi_0}, \ket{\psi_1}, \delta)$ is equal to $\min_U(C(U))$ such that 
        \begin{equation}
            \frac{\abs{\bra{\psi_1}U\ket{\psi_0} + \bra{\psi_0}U\ket{\psi_1}}}{2} \geq \delta.
        \end{equation}
\end{definition}

\begin{definition}[Distinguishability complexity proxy]
    \label{def:distinguishability-complexity}
    Let $0 \leq \delta \leq 1$. The distinguishability complexity $C_D(\ket{\psi_0}, \ket{\psi_1}, \delta)$ is equal to $\min_U(C(U))$ such that 
    \begin{equation}
        \frac{\abs{\bra{\psi_0}U\ket{\psi_0} - \bra{\psi_1}U\ket{\psi_1}}}{2} \geq \delta.
    \end{equation}
\end{definition}

\noindent It was shown in~\cite{aaronson2020hardness} that these proxy metrics
are within a constant factor $\mathcal{O}(1)$ of the true interference and
distinguishability complexities~\footnote{Throughout this paper, we use the
somewhat more abbreviated naming convention of ``distinguishability complexity
proxy'' and ``interference complexity proxy'' to refer to the concepts of
``distinguishability proxy'' and ``interference proxy'', respectively.}.

\begin{definition}[Branch factor]
    \label{def:branch-factor}
    Let $0 \leq \delta \leq 1$. The branch factor~\footnote{This is called
    ``branchiness'' in~\cite{taylor2023wavefunction}.} between states
    $\ket{\psi_0}$ and $\ket{\psi_1}$ is defined as
    \begin{equation}
        B(\ket{\psi_0}, \ket{\psi_1}, \delta) := C_I(\ket{\psi_0}, \ket{\psi_1}, \delta) - C_D(\ket{\psi_0}, \ket{\psi_1}, \delta).
    \end{equation}
\end{definition}
We consider a branch factor ``good'' when the interference proxy is
significantly greater than the distinguishability proxy, i.e. $C_I(\ket{\psi_0},
\ket{\psi_1}, \delta) \gg C_D(\ket{\psi_0}, \ket{\psi_1}, \delta)$ for some
choice of $\ket{\psi_0}$, $\ket{\psi_1}$, and $\delta$.

To illustrate the idea behind these definitions, consider the decomposition
$\ket{\psi} = \frac{\ket{a} + \ket{b}}{\sqrt{2}}$ into two branches $\ket{a}$
and $\ket{b}$. A calculation shows that observing different measurement outcome
probabilities for the pure state density matrix $\ket{\psi}\bra{\psi}$ and the
classical mixture $\frac{\ket{a}\bra{a} + \ket{b}\bra{b}}{2}$ is as hard as
swapping $\ket{a}$ and $\ket{b}$~\cite{aaronson2020hardness}. This hardness is
given by the interference complexity and is a measure of how many gates you need
to obtain relative phase information. For good branch decompositions, we want
this interference complexity to be high (so the state behaves like a classical
mixture as far as the measurement outcomes are concerned), but this is not
enough. It should also be relatively easy to tell the difference between the two
different branches. The complexity required for this is given by the
distinguishability complexity.

Some examples of states with good branch factors, as discussed
in~\cite{taylor2023wavefunction}, are
\begin{itemize}
    \item GHZ state: Let $n\geq 1$ be a positive integer and let $\ket{\psi_0} =
    \ket{0^n}$ and $\ket{\psi_1} = \ket{1^n}$. Then, for $\delta = 1$, we can
    exactly compute the interference and distinguishability proxy. For
    interference, we can use two-qubit gates $X\otimes X$ to flip the bits two
    at a time to swap $\ket{0^n}$ with $\ket{1^n}$. If $n$ is even, we need
    $n/2$ two-qubit gates, while in the case that $n$ is odd, we need $(n-1)/2$
    two-qubit gates and one single-qubit gate. So, the interference proxy is $n$
    regardless of the parity of $n$. For the distinguishability proxy, we need a
    circuit that maps $\ket{0^n}$ to itself, but $\ket{1^n}$ to $-\ket{1^n}$. We
    can do this by applying one $Z$-gate on the first qubit. So
    \begin{equation}
        \label{eq:GHZ-bf}
        C_I = n, \quad C_D = 1, \quad \text{and} \quad B = n - 1.
    \end{equation}
    \item Product state and Haar random state: for $n$-qubit states, let
    $\ket{\psi_0} = \ket{0^n}$ and $\ket{\psi_1} = \ket{\upsilon}$, where
    $\ket{\upsilon}$ is a Haar random state. Then for
    $\delta=1$~\cite{knill1995approximation}
    \begin{equation}
        \label{eq:RU-bf}
        C_I \geq  4^n/9  - n/3 -1/9, \quad C_D \leq n, \quad \text{and} \quad B \geq 4^n/9  - 4n/3 -1/9,
    \end{equation}
    with high probability (probability of measuring $0^n$ in a Haar random
    quantum state is $O(\exp(-n))$).
    \item Two random states $n$-qubit states produced by circuits with depth
    $D_0$ and $D_1$ respectively. It can be shown~\cite{taylor2023wavefunction}
    that
    \begin{equation}
        \label{eq:two-random-BF}
        C_I = \mathcal{O}((D_0+D_1)n) \quad \text{and} \quad C_D = \mathcal{O}(\min(D_0, D_1)n)
    \end{equation}
    \item Product state and Dicke state: for $n$-qubit states, let $\ket{\psi_0}
    = \ket{0^n}$ and $\ket{\psi_1} = D(n,n/2)$ where $D(n,k)$ is a Dicke state:
    an equal superposition over all bitstrings of size $n$ that have Hamming
    weight $k$. Then, for $\delta=1$, assuming the asymptotic upper bound of the
    circuit complexity in~\cite{bartschi2019deterministic} is also a lower bound
    \begin{equation}\label{eq:DS-bf}
        C_I \geq \Omega(n^2), \quad C_D = O(1), \quad \text{and} \quad B = \Omega(n^2),
    \end{equation}
    where $B=\Omega(n^2)$ is shorthand for the statement that there is a $C>0$
    such that for all $n\geq n_0$ we have $B\geq Cn^2$ for some integer $n_0$.
    The distinguishability proxy is $O(1)$ with high probability: measuring a
    random qubit gives us a 1 with probability $\sim 1/2$, so measuring a
    constant number of qubits, one can get a success probability as high as one
    wants.
\end{itemize}

Examples of ``bad'' branch factors are those that have the reverse relationship,
where the distinguishability proxy is significantly greater than the
interference proxy, i.e., $C_D(\ket{\psi_0}, \ket{\psi_1}, \delta) \gg
C_I(\ket{\psi_0}, \ket{\psi_1}, \delta)$ for some choice of $\ket{\psi_0}$,
$\ket{\psi_1}$, and $\delta$. As considered in~\cite{taylor2023wavefunction},
error-correcting codes have bad branch factors. Making it hard for the
environment to get information from your quantum system (i.e., high
distinguishability) makes it less likely for your system to decohere.

In principle, one could compute the branch factor for any quantum system in a
superposition of (preferred) pointer states. This includes large quantum systems
like molecules, but in practice, it can be very challenging to compute the
branch factor exactly.

Section~\ref{sec:experimental-violations} describes experiments where GHZ states
are used to increase branch size. GHZ state friends also increase the friend's
particle number and the size of the Hilbert space, which are other interesting
candidates for observerness. Connecting the branch factor metric of observerness
to other proposals (Section~\ref{sec:other-observer-measures}) is an open avenue
for future work.

By running the EWFS with friends with progressively larger branch factors, we
test LF violations for increasingly meaningful classes of observers. What could
the outcomes of such a program be?
\begin{enumerate}
    \item We could observe violations up to systems with the same branch factors
    as humans or even with humans as friends. Each result at each scale would
    provide more evidence that LF is indeed violated by reality.
    \item We could observe some threshold branch factor where the LF violations
    stop. One could interpret this as some ``collapse model" where friends above
    that threshold count as observers. Thus, one could argue that LF is not
    violated by reality. Notably, this would require some extension to textbook
    quantum mechanics and provide experimental evidence for a specific
    definition of an observer.
\end{enumerate}

Of course, pursuing this program involves significant engineering challenges in
reliably preparing increasing branch factors. Importantly, it also requires us
to consider how to handle experiments where branch factors are produced under
noisy experimental conditions. We say more about handling these noisy conditions
in Section~\ref{sec:noisy-branches} and Section~\ref{sec:ghz-friends}.

In the rest of the paper, we only consider the value $\delta=1$ in the
definition of branch factor. This would mean that we are only interested in
perfect interference and distinguishability. The main reason for this is the
fact that the complexities are easier to analyze. Using $\delta=1$, the values
for the complexities become upper bounds, this is however not necessarily true
for the branch factor.

%%%%%%%%%%%%%%%%%%%%%%%%%%%%%%%%%%%%%%%%%%%%%%%%%%%%%%
\section{Experimental Local Friendliness violations on quantum computers}
\label{sec:experimental-violations}
%%%%%%%%%%%%%%%%%%%%%%%%%%%%%%%%%%%%%%%%%%%%%%%%%%%%%%

In this section, we empirically demonstrate LF inequality violations using
Extended Wigner's Friend Scenarios for increasing branch factors of the friend
system using (i) a noiseless quantum processor simulator, (ii) a simulator with
noise models, (iii), and real quantum hardware devices.

The friend system we use in this section is the GHZ state. To experimentally
demonstrate these violations, we use approaches to validate branch factors
produced under noise and error mitigation approaches to deal with measurement
noise.

While one can consider scenarios where all local friendliness inequalities are
violated, a violation of any one of inequalities is sufficient to illustrate
that local friendliness has been violated. We, therefore, focus on violating the
semi-Brukner inequality from Equation~\eqref{eq:semi-brukner-lf-inequality} in
the following sections. The reason for focusing on this inequality is its
relative simplicity. Specifically, notice that
Equation~\eqref{eq:semi-brukner-lf-inequality} lacks a $B_1$ observable,
eliminating the need to consider a PEEK setting for Bob. This simplification
allows us to simplify Bob's friend (Debbie) to just a single qubit, as removing
PEEK eliminates the need for Bob to look at Debbie's qubit register.\footnote{We
thank Eric Cavalcanti, Howard Wiseman, and Nora Tischler for pointing this out
to us.} Removing this setting configuration from our circuit allows us to reduce
the size of our circuit by approximately a factor of two. This enables us to
scale the circuit and consider larger friend sizes on simulators and hardware
devices that would have otherwise been more difficult to obtain. As shown in
Section~\ref{sec:semi-brukner-no-debbie}, the semi-Brukner inequality still
holds in this case.

To compute the expectation values for the semi-Brukner inequality, we have to
infer what branch the friends are in using a single measurement. Since the
friend system has low distinguishability proxy for the systems we consider, this
is possible with high probability. The following subsections describe how we do
this for the GHZ and controlled-random unitary friend systems and plot the
empirically obtained LF inequality violations.

\begin{figure}[h]
    \centering
    \includegraphics[scale=1.0]{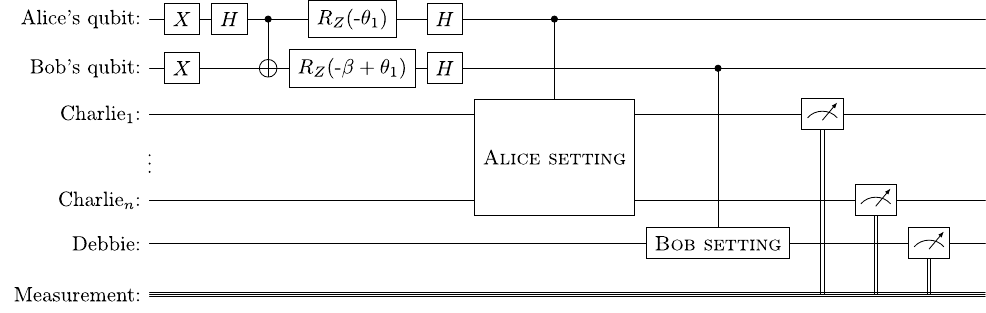}
    \caption{\textbf{Circuit depiction of the PEEK-PEEK setting for the extended
    Wigner's Friend scenario.} Alice and Bob begin by preparing a bipartite
    state as defined in Equation~\eqref{eq:ewfs-bipartite-state}. Alice then
    performs her measurement setting on Charlie's qubit(s); likewise, Bob
    performs his measurement on Debbie's (single) qubit. The settings performed
    by Alice and Bob is either PEEK, REVERSE-1, or REVERSE-2. In this particular
    depiction of the circuit, we assume that Alice and Bob are performing the
    PEEK setting. Finally, the system qubits of Charlie and Debbie are
    measured.}
    \label{fig:ewfs-circuit}
\end{figure}

The general form of the EWFS circuit is shown in Figure~\ref{fig:ewfs-circuit}.
Alice and Bob's measurement settings (e.g., PEEK, REVERSE-1, or REVERSE-2 as
defined in Section~\ref{sec:qm-violation-three-setting-two-party}) cause the
controlled operations to be labeled as ALICE SETTING and BOB SETTING. The
specific gates in these controlled operations depend on what type of friend is
instantiated. Using a CNOT ladder, as in Figure~\ref{fig:cnot-ladder}, causes
the friend system to be in a GHZ state for which we know the branch factor
increases linearly in the friend's size, c.f. Equation~\eqref{eq:GHZ-bf}. If we
use a controlled-random unitary, the branch factor will increase exponentially
in the friend system size, c.f. Equation~\eqref{eq:RU-bf}. However, we need
several gates that grow exponentially in system size to implement such a random
unitary.

\begin{figure}[h]
    \centering
    \begin{quantikz}[thin lines, row sep=0.2cm, column sep=0.2cm]
        \ket{0}_1 & 
        \qw & 
        \ctrl{1} &
        \qw &
        \qw &
        \qw &
        \qw \\
        \ket{0}_2 & 
        \qw & 
        \targ{} & 
        \ctrl{1} &
        \qw &
        \qw &
        \qw \\
        \ket{0}_3 & 
        \qw & 
        \qw &
        \targ{} &
        \qw &
        \qw &
        \qw \\
        \vdots \\ 
        \ket{0}_{n-1} &
        \qw & 
        \qw &
        \qw &
        \qw &
        \ctrl{1} &
        \qw \\
        \ket{0}_n & 
        \qw & 
        \qw &
        \qw &
        \qw &
        \targ{} &
        \qw 
    \end{quantikz}
    \caption{\textbf{CNOT ladder for GHZ friends in EWFS.}  An $n$-qubit CNOT
    ladder used by Charlie in the \textsc{ALICE SETTING} circuit of
    Figure~\ref{fig:ewfs-circuit}.}
    \label{fig:cnot-ladder}
\end{figure}

Using the circuit description of the EWFS scenario, one can quickly compute that
$\expval{ A_i} = \expval{B_i} = 0$ for all $i=1,2,3$ and that $\expval{A_i B_j}
= -\cos(\beta_j-\theta_i)$. This allows us to find the optimal angles for
maximal violation of the semi-Brukner inequality
\begin{equation}
    \cos(\beta_2-\theta_1) - \cos(\beta_3-\theta_1)
    +\cos(\beta_2-\theta_3)+\cos(\beta_3-\theta_3) - 2\leq 0.
\end{equation}

\noindent We give the maximum inequality violations under optimal choices of
angles in the right column of Table~\ref{tab:ewfs-violations}.

%%%%%%%%%%%%%%%%%%%%%%%%%%%%%%%%%%%%%%%%%%%%%%%%%%%%%%
\subsection{Noisy Friends: preparing branch factors under noise}
\label{sec:noisy-branches}
%%%%%%%%%%%%%%%%%%%%%%%%%%%%%%%%%%%%%%%%%%%%%%%%%%%%%%

In this work, our EWFS are prepared on quantum computers, which have errors in
their operation. Even if one is not using a quantum computer, any experimental
setup will have noise in the prepared friend states. Thus, we need a model for
dealing with LF violations where we could have errors in the branch factors of
the prepared friend states.

One approach is to characterize a fidelity bound on the state of Charlie in the
middle of the scenario. For example, say we aim for Charlie to be in the friend
state $\ket{\psi}:= \frac{1}{\sqrt{2}}\ket{\psi_0}+\ket{\psi_1}$ after they have
measured their system. Errors in that preparation mean that Charlie is instead
in a state represented by the density matrix $\rho$. Then the fidelity
$\mathcal{F}(\ket{\psi}, \rho) = \bra{\psi}\rho\ket{\psi}$ gives a lower bound
$q$ on the probability that the state $\ket{\psi}$ has been prepared with our
target branch factor. 

Our experiments must show that we have prepared the intended branch factor and
violated local friendliness with those same states. We need to avoid a scenario
where the incorrect states dominate the expectation values and contribute to the
local friendliness violation. Consider a proposed experimental violation of the
semi-Brukner inequality.
\begin{equation}
    X:= -\expval{A_1B_2}+\expval{A_1B_3}-\expval{A_3B_2}-\expval{A_3B_3}\ge 2
\end{equation}
When $X \ge 2$, we have an LF violation. To bound the value of $X$ as a function
of the probability $q$ that the data going into the expectation is valid, we
need to consider the worst-case scenario for the expectation values given the
bounds and the probability $1 - q$ of invalid data. Each random variable $A_i$
and $B_j$ is bounded between -1 and 1. If there is a probability $q$ that the
data is invalid, the invalid data might skew the expectation values. For invalid
data, we assume the worst-case scenario where the invalid data maximizes or
minimizes each term in $X$. The effective expectation value can be modeled as
a mixture
\begin{equation}
    \expval{A_iB_j} = q \expval{A_iB_j}^{\text{valid}} + (1-q) \expval{A_iB_j}^{\text{invalid}}
\end{equation}
Similarly, for the other expectation values. Let $\expval{\widetilde{A_iB_j}}$
be the measured expectation value. This means that 
\begin{equation}
   \expval{A_iB_j}^{\text{valid}} = \frac{\expval{\widetilde{A_iB_j}} - (1-q) \expval{A_iB_j}^{\text{invalid}}}{q}
\end{equation}
We then set $\expval{A_iB_j}^{\text{invalid}}$ to whatever is the worst case for
that term in $X$. For the positive term $\expval{A_1 B_3}$ we have 
\begin{equation}
   \expval{A_1 B_3}^{\text{valid}} \ge \frac{\expval{\widetilde{A_1 B_3}} - 2(1-q)}{q}
\end{equation}
For the negative terms, we have
\begin{equation}
   -\expval{A_iB_j}^{\text{valid}} \ge 
   \frac{2(q-1)-\expval{\widetilde{A_iB_j}}}{q}
\end{equation}
and similarly for the other terms. This gives 
\begin{equation}
    \begin{aligned}
        X^{\text{valid}}&:= -\expval{A_1B_2}^{\text{valid}}+\expval{A_1B_3}^{\text{valid}}-\expval{A_3B_2}^{\text{valid}}-\expval{A_3B_3}^{\text{valid}} \\
        X^{\text{valid}}&\ge
        \frac{1}{q} \left[
        \tilde{X}
        +8(q-1)
        \right]
    \end{aligned}
\end{equation}

The value of $X^{\text{valid}}$ is what matters to show the violation under
noise. We must have $X^{\text{valid}} \ge 2$ or similarly, the measured value
\begin{equation}
    \tilde{X} \ge 8 - 6q.
\end{equation}
The maximum value for $\tilde{X}$ in the Semi-Brukner inequality is 2.82843 (See
Table~\ref{tab:ewfs-violations}). Solving for the minimum fidelity that supports
this gives $93.66\%$. Thus, we don't need perfect friend state preparation to
see violations. If our friend state is prepared with greater than $93.66\%$,
then we can be confident that our LF violations hold at our target branch
factor.

Now, we consider how to characterize $q$ from a real experiment. One approach is
to do state tomography and measure it directly. Unfortunately, this becomes
experimentally difficult for larger friend sizes. Instead, we propose
extrapolating from a depolarizing model for the quantum circuit. 

\emph{Depolarizing Case:} Assume a depolarizing noise model where we apply the
depolarizing channel after each single-qubit gate in the circuit with parameter
$p_1$ and with parameter $p_2$ after each two-qubit gate, i.e.,
$\mathcal{E}(\rho) = (1-p_i)\rho + p_iI/2^n$ for $n$-qubit quantum state $\rho$.
Then it is straightforward to check that if our circuit with $n$ single-qubit
gates and $m$ two-qubit gates are supposed to prepare the quantum state $\rho =
\ket{\psi}\bra{\psi}$, the noisy density matrix will be $\rho' = (1-p_1)^n
(1-p_2)^m\rho + (1-(1-p_1)^n (1-p_2)^m) I/2^n$. Since the above observables are
zero when evaluated at the maximally mixed state, then by linearity of the
expectation value, the expected values of the observables for the noisy density
matrix $\rho'$ will be equal to the ideal one multiplied by $(1-p_1)^n
(1-p_2)^m$. Thus, we'd need $\tilde{X} \ge 2/[(1-p_1)^n (1-p_2)^m]$ to show a
violation. Again, using the max value of 2.82843 for $\tilde{X}$, we see that
the minimum single-qubit and two-qubit fidelities should satisfy $(1-p_1)^n
(1-p_2)^m\ge 2/2.82843 \approx 0.707$. Usually, the two-qubit gate fidelity
$1-p_2$ is much lower than the single-qubit gate fidelity, and the preparation
of the friend size mostly dominates the number of two-qubit gates. Linearly
increasing friend size increases the two-qubit gates linearly, assuming a
chain-like layout for the friend system. 

For a circuit with $10$ layers of gates (where each layer represents one step of
parallel gate operations in the circuit depth), assuming $p_1 = p_2/10$ and
$n=10m$ (single-qubit gates being an order of magnitude higher fidelity and
friend system preparation requiring an order of magnitude more single-qubit
gates), this would need a two-qubit gate error rate of at most $\sim 1.7\%$.
This is not an infeasible error rate for near-term quantum computers, but to
scale to a circuit with 100 layers of depth, that error rate also drops by an
order of magnitude. If depolarizing noise is a good error model; each order of
magnitude decrease in error-rate should allow one to push to another order of
magnitude in friend size and branch factor. Fault-tolerant quantum computers
will be critical here.

%%%%%%%%%%%%%%%%%%%%%%%%%%%%%%%%%%%%%%%%%%%%%%%%%%%%%%%%%%%%%%%%%%%%%%%%%%%%%%%%%%%%%%%%%%%%%%
\subsubsection{Other approaches to validate branch factors}
\label{sec:validate-branch-factors}
%%%%%%%%%%%%%%%%%%%%%%%%%%%%%%%%%%%%%%%%%%%%%%%%%%%%%%%%%%%%%%%%%%%%%%%%%%%%%%%%%%%%%%%%%%%%%%
The approach described above bootstraps up to estimated fidelities from
characterizing modular components. One might prefer a systematic witness for
branch factor that doesn't require such extrapolation. 

Consider the case where friend states are given by two Haar random states whose
interference and distinguishability proxies are then given
by~\eqref{eq:two-random-BF}. We know there is a signature for random quantum
states given by heavy output distributions whose statistics can act as a proxy
for having prepared random quantum states~\cite{arute2019quantum}. This would
give a protocol for validated EWFS experiments where one randomly interleaves
the true experiment, gathering Alice and Bob's statistics, or validations where
computational basis measurement statistics are taken on Charlie's and/or
Debbie's qubits. The measurements from the validation runs can be verified using
cross-entropy benchmarking following~\cite{arute2019quantum}.

A downside of this direction is that this verification is computationally
intensive, requiring classical simulation of the circuits to do the
cross-entropy benchmarking. Thus, this approach is likely limited to $\le100$
qubits. Instead, a validation-based collision counting of random quantum states
would not require much classical computing but does require a large amount of
sampling~\cite{mari2023counting}. Another alternative is to consider ``peaked''
random circuits proposed for efficient validation~\cite{aaronson2024verifiable}.
However, more work needs to be done to study the preparation and branch factors
of superpositions of these states.\footnote{We thank Andrea Mari for pointing
out these recent works on validation to us.}

More broadly, we would prefer classically efficient verifications that the
quantum states of sufficient branch factor have been produced.
In~\cite{reichardt2013classical}, the authors introduce a protocol for
classically verifying a quantum system's dynamics using CHSH tests. More
examples of ``self-testing'' to verify the production of specific quantum states
are given in this survey~\cite{vsupic2020self}. For example,
in~\cite{chabaud2021efficient} a protocol is given to use single-mode Gaussian
measurements to verify a class of continuous variable quantum states, including
Boson Sampling states. This class of states could be used to design friends who
have high branch factors but also efficient state verification. We leave it to
future work to directly link the self-testing literature to validating branch
factors, but note that this direction looks promising for scaling up
validations.

%%%%%%%%%%%%%%%%%%%%%%%%%%%%%%%%%%%%%%%%%%%%%%%%%%%%%%
\subsection{Noisy Measurements: inferring observed outcomes using majority vote}
\label{sec:ghz-friends}
\begin{figure}[!htpb]
    \centering
    \includegraphics[scale=0.4]{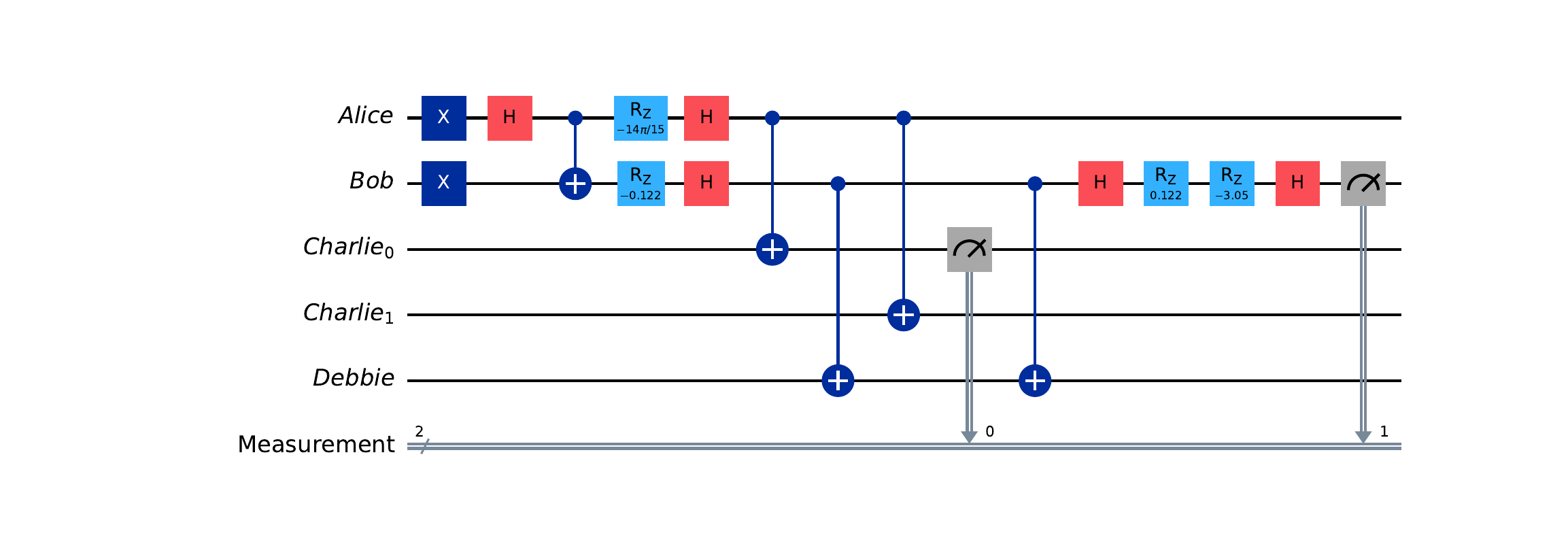}
    \caption{\textbf{Example EWFS circuit with PEEK and REVERSE-1 settings.} An
    example of the EWFS circuit with GHZ friends using the random strategy,
    where the friend Charlie consists of two qubits. Alice uses a PEEK setting
    in this example circuit, while Bob uses the REVERSE-1 setting.}
    \label{fig:ewfs-peek-reverse-1}
\end{figure}

\begin{figure}[!htpb]
    \centering
    \includegraphics[scale=0.4]{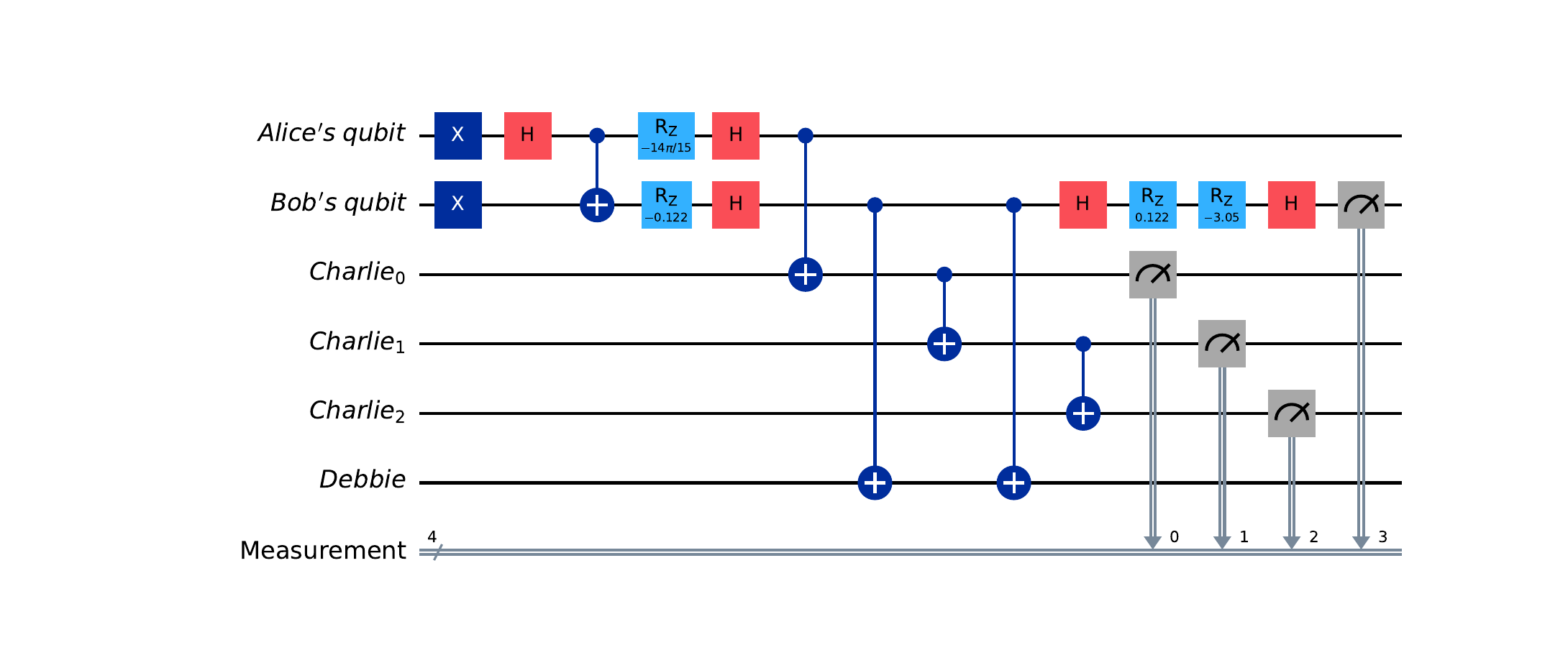}
    \caption{\textbf{EWFS circuit where Alice measures all qubits of Charlie.}
    An example of the EWFS circuit with GHZ friends using the majority vote
    strategy, where the friend Charlie consists of three qubits. Alice uses a
    PEEK setting in this example circuit, while Bob uses the REVERSE-1 setting.
    Note that Alice is measuring all the qubits of her friend in this case.}
    \label{fig:ewfs-circuit-majority-vote-peek-reverse-1}
\end{figure}

%%%%%%%%%%%%%%%%%%%%%%%%%%%%%%%%%%%%%%%%%%%%%%%%%%%%%%
After measurement, there are many ways to infer the friend's outcome in the PEEK
setting. A simple way would be to randomly pick a qubit from the friend system
and measure it. In the noiseless case, this would work perfectly to determine
the outcome of the GHZ state. An example circuit of EWFS using this approach is
in Figure~\ref{fig:ewfs-peek-reverse-1}. However, this will not work well in the
presence of measurement noise since bits might get flipped by measurement
errors. A better approach would be to measure multiple qubits and decide which
outcome you have measured based on a majority vote.
Figure~\ref{fig:ewfs-circuit-majority-vote-peek-reverse-1} shows an example
where we measure all the qubits of the friend system in the PEEK setting. The
observables we are interested in assign a value of $1$ to one outcome and a
value of $-1$ to the other.

For the case of GHZ friends, when measuring the all-zero state in the
computational basis, we want to assign a value of $1$, and when measuring the
all-ones state, we assign a value of $-1$. In the presence of noise, other
bitstrings can be measured as well. We use the majority vote to determine the
value of the other bitstrings. If $n$ is the size of the friend system, which we
assume to be odd, the measurement consists of the following positive
semidefinite matrices
\begin{equation}
    F_0 = \sum_{H(x) < n/2} \ketbra{x}{x}
    \quad \text{and} \quad 
    F_1 = \sum_{H(x) > n/2} \ketbra{x}{x}
\end{equation}
where $H(x)$ is the Hamming weight of the $n$-bit string $x$. The observable for
Alice is then simply
\begin{equation}
    A = F_0 - F_1.
\end{equation}
The observable $B$ for Bob is defined similarly.

\begin{figure}[!htpb]
    \centering
    \includegraphics[scale=0.4]{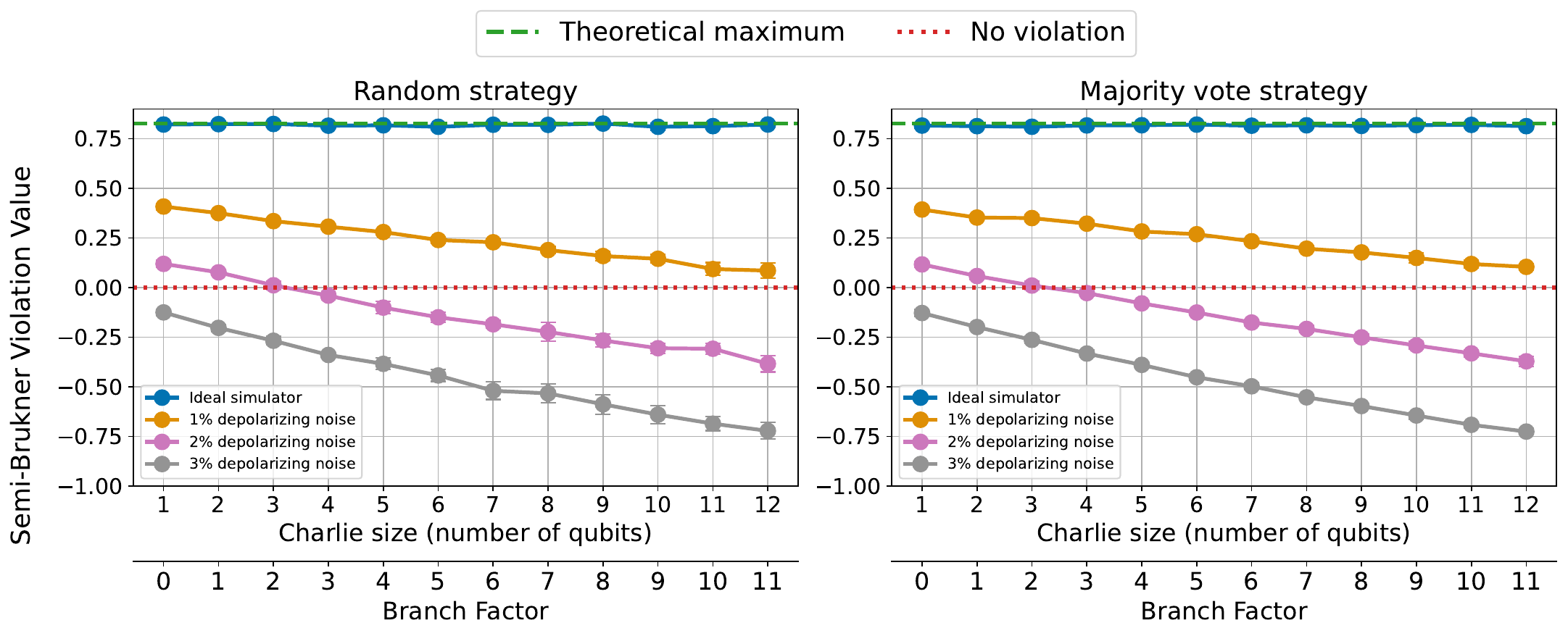}
    \caption{\textbf{A comparison between the ``random'' and ``majority vote''
    strategies for EWFS.} Each plot considers how each strategy performs for
    progressively increasing depolarizing noise levels with fixed $1\%$ readout
    error on all qubits. Each data point is run using 10000 shots and averaged
    over 10 trials. The top x-axis ranges over the number of qubits in the
    quantum system size of Charlie, while the bottom x-axis shows the
    corresponding branch factor. One can see there is minimal difference between
    the two approaches, except that the majority vote approach has less
    variance.}
    \label{fig:ghz-strategy-simulator-comparison}
\end{figure}

In the ideal scenario, when there is no noise, the violations of the LF
inequalities should match the approach where we infer the friend's observed
outcome by measuring a random qubit of the friend. However, as we introduce
noise, while the magnitude of the LF violations will be the same on average,
there will be more variation in the random approach. We demonstrate this by
comparing the simulated violations using these two approaches. The results of
running the EWFS, where we measure just one qubit of the friend system vs. the
use of the majority vote observable for GHZ friends of increasing qubit sizes on
quantum simulators using depolarizing noise and readout error are depicted in
Figure~\ref{fig:ghz-strategy-simulator-comparison}.

This majority vote approach is one form of readout error mitigation, and future
work can apply more of these methods. For example, stabilizer checks on the
friend's unitary operation can be used for error detection and post-selection. A
bit-flip error in constructing the GHZ state for Charlie still results in states
with the same branch factor. One can consider this bit flip like a relabeling of
the ``logical" GHZ states from, for example, $|000\rangle = |O\rangle_{L};
|111\rangle = |1\rangle_{L}$ to $|001\rangle = |O\rangle_{L}; |110\rangle =
|1\rangle_{L}$. However, Alice needs to know on which qubit the error occurred
to not incorrectly infer what branch Charlie is in. In this case, tracking an
error on the third qubit means Alice can assign the inverse of its measurement.
As quantum processors become more sophisticated, the full suite of error
mitigation and correction techniques can be deployed to characterize the target
friend states with high fidelity.

%%%%%%%%%%%%%%%%%%%%%%%%%%%%%%%%%%%%%%%%%%%%%%%%%%%%%%%%%%%%%%%
\subsection{Experimental Results: LF Violations using GHZ friend states on quantum computers}
\label{sec:experiments}
%%%%%%%%%%%%%%%%%%%%%%%%%%%%%%%%%%%%%%%%%%%%%%%%%%%%%%%%%%%%%%%
\begin{figure}[!htpb]
    \centering
    \includegraphics[scale=0.5]{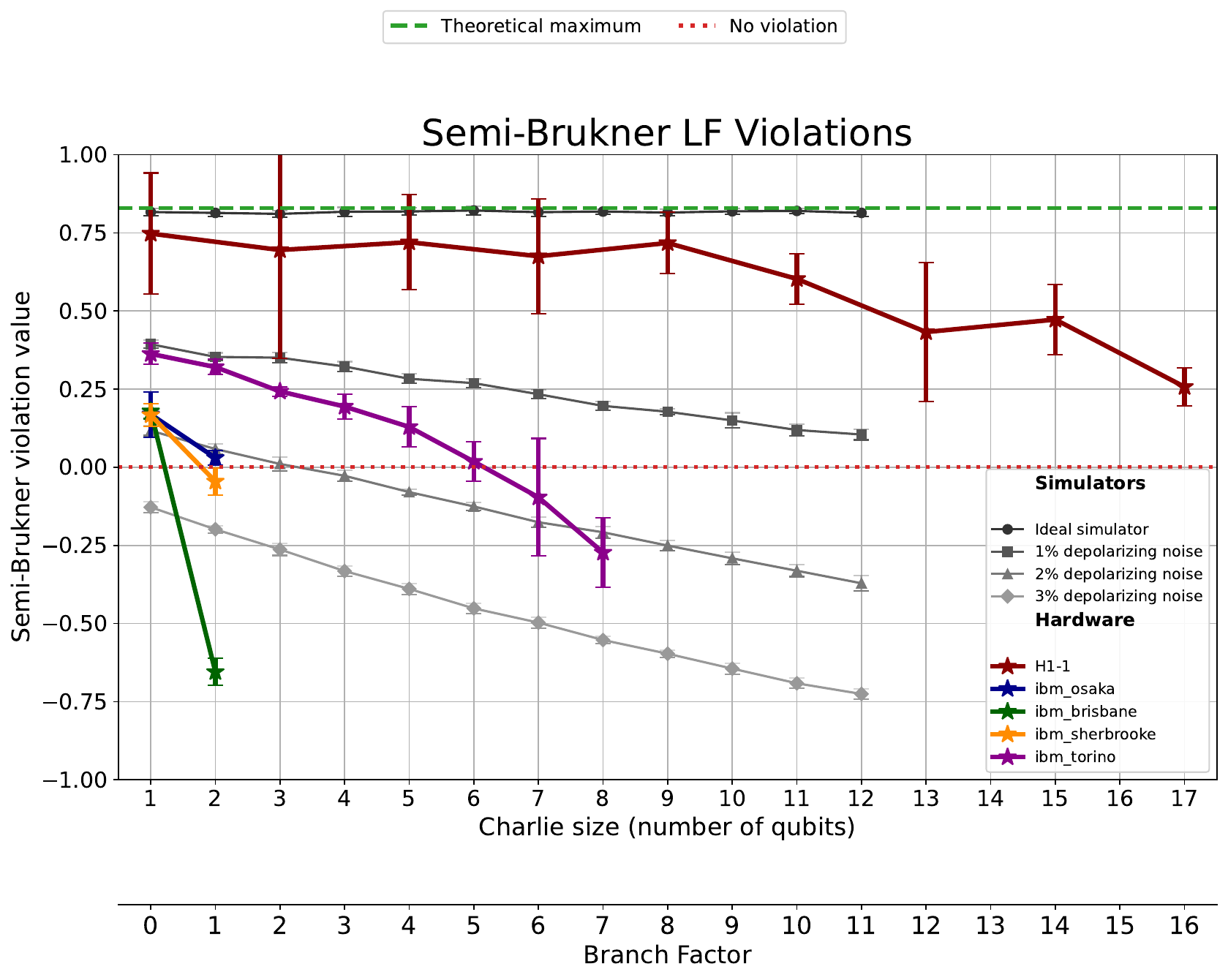}
    \caption{\textbf{A comparison between simulator, noisy emulator, and
    hardware for majority vote EWFS.} The gray-scaled lines show how increasing
    the depolarizing noise reduces the maximum friend size for which a violation
    occurs. The lines in color show the \texttt{ibm\_osaka},
    \texttt{ibm\_sherbrooke}, \texttt{ibm\_torino} IBM hardware devices and the
    H1-1 Quantinuum hardware device. Note that the only IBM hardware device to
    obtain violations beyond branch factor 0 is \texttt{ibm\_torino}, showing a
    violation at branch factor 4, while on the H1-1 device, we obtain a
    violation at branch factor 16. We transpile the circuit before running for
    the specific coupling map topology for the emulated and real hardware.
    Further details on the transpilation are provided in
    Appendix~\ref{app:running-ewfs-on-quantum-hardware}. For each line in the
    plot, the top x-axis ranges over the number of qubits in the quantum system
    size of Charlie, while the bottom x-axis shows the corresponding branch
    factor. All IBM data points are run with 10000 shots over 10 trials. For the
    H1-1 device, we ran each experiment with 200 shots and 4 trials, except for
    the experiment of size 17, which we ran for 7 trials. Error bars are 3
    standard deviations.}
    \label{fig:ghz-majority-vote-simulator-fakehardware-hardware-comparison}
\end{figure}

Using the majority vote approach for measurement described previously, we
perform experiments on IBM superconducting quantum computers to measure
violations. Results from noisy model simulations and hardware experiments are
plotted in
Figure~\ref{fig:ghz-majority-vote-simulator-fakehardware-hardware-comparison}.
These experiments use the majority vote implementation of EWFS from
Section~\ref{sec:ghz-friends}. In these plots, the y-axis is the value of the
Semi-Brukner violation on the left-hand side of the inequality in
Equation~\eqref{eq:semi-brukner-lf-inequality}. Thus, we have a violation when
this value is above 0, as indicated by the red dotted line. The green dotted
lines indicate the maximum achievable theoretical value of 0.82843 (see
Section~\ref{sec:experimental-violations}). We plot these violations for
experiments where the friend is prepared in a GHZ state of increasing qubit
number (bottom x-axis) corresponding to branch factors indicated on the top
x-axis. The point where the violation value crosses below the dotted red line is
the maximum branch factor for which we can show violations on that backend.

The leftmost plot shows results from a simulated backend running perfect
simulations or quantum circuits at 1\%, 2\%, and 3\% per gate depolarizing noise
rates. Perfect simulations have no limit to the size of violations we expect to
see, corresponding to the predictions of textbook quantum mechanics. For each
noise rate, we see violations decrease such that a 2\% per gate noise rate would
only support violations to branch factor 9, and a 3\% per gate noise rate
supports a validation up to branch factor 4. These depolarizing simulations do
not include any specific gate set or qubit topology compilations. Thus, we
expect them to be optimistic predictions for real experiments where that
compilation adds additional overhead.

The middle plot shows violations on various hardware emulators, including more
sophisticated noise models based on several superconducting and one ion trap
emulator.

The rightmost plot shows violations obtained directly on quantum hardware: IBM's
superconducting quantum processors Osaka, \texttt{ibm\_sherbrooke}, and
\texttt{ibm\_torino}. While \texttt{ibm\_osaka} and \texttt{ibm\_sherbrooke} do
not show violations beyond branch factor 0, the \texttt{ibm\_torino} processor
supports violations up to branch factor 4. Importantly, fidelity is not the only
driver of performance here. The \texttt{ibm\_torino} processor's native gates
are better suited to our circuit, so it also has much lower gate counts,
improving performance. On the H1-1 Quantinuum device~\cite{quantinuum_h1}, we
obtained a violation at the branch factor as high as 16.

%%%%%%%%%%%%%%%%%%%%%%%%%%%%%%%%%%%%%%%%%%%%%%%%%%%%%%%%%%%%%%%%%%%%%%%%%%%%%%
\subsubsection{Validation of branch factor preparation}
\label{sec:validation-branch-factor-prep}
%%%%%%%%%%%%%%%%%%%%%%%%%%%%%%%%%%%%%%%%%%%%%%%%%%%%%%%%%%%%%%%%%%%%%%%%%%%%%%

\begin{figure}[!htbp]
    \centering
    \begin{subfigure}[b]{0.44\textwidth}
        \centering
        \includegraphics[width=\textwidth]{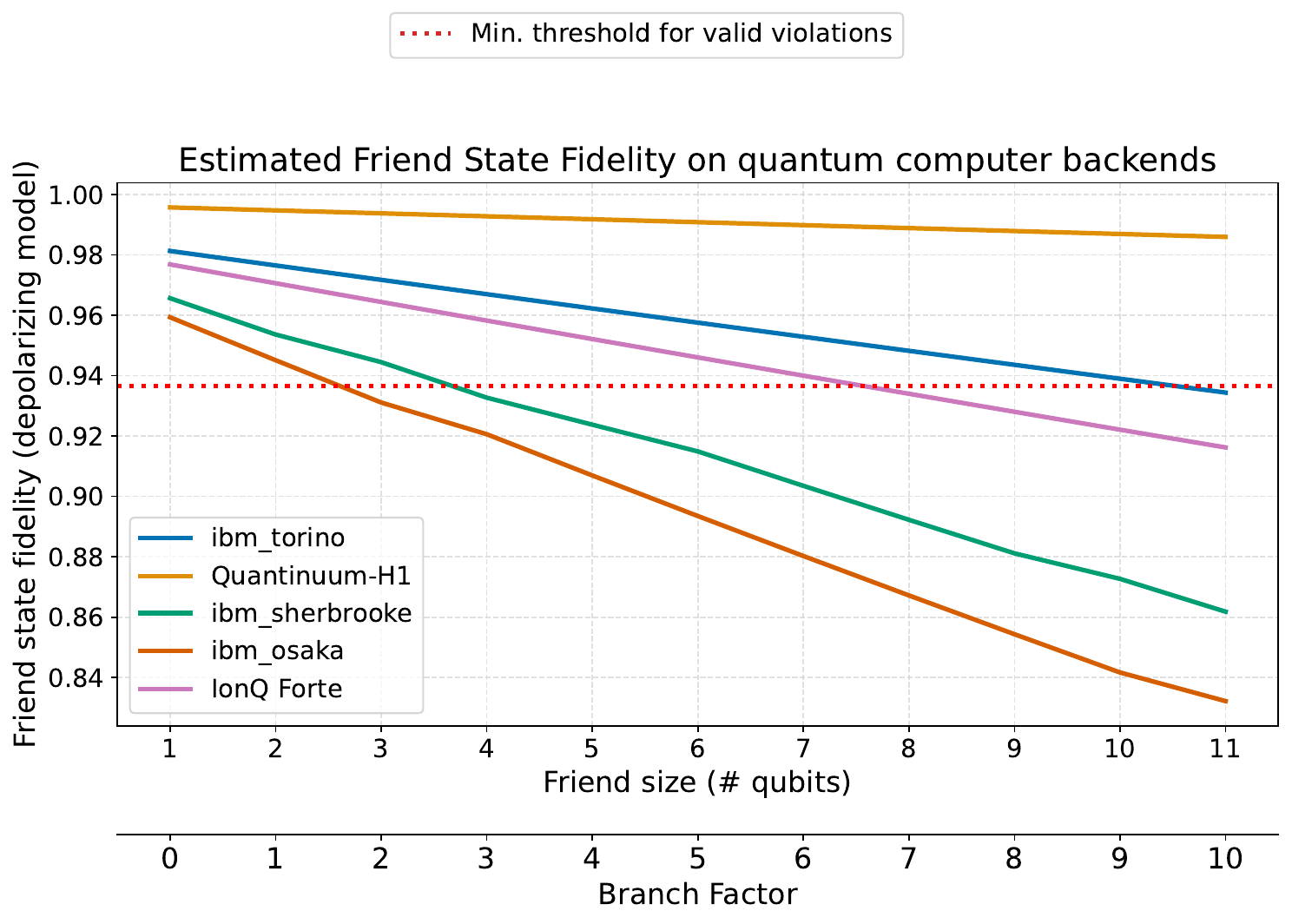}
        \label{fig:subfig1}
    \end{subfigure}
    \hspace{0.3cm}
    \begin{subfigure}[b]{0.44\textwidth}
        \centering
        \includegraphics[width=\textwidth]{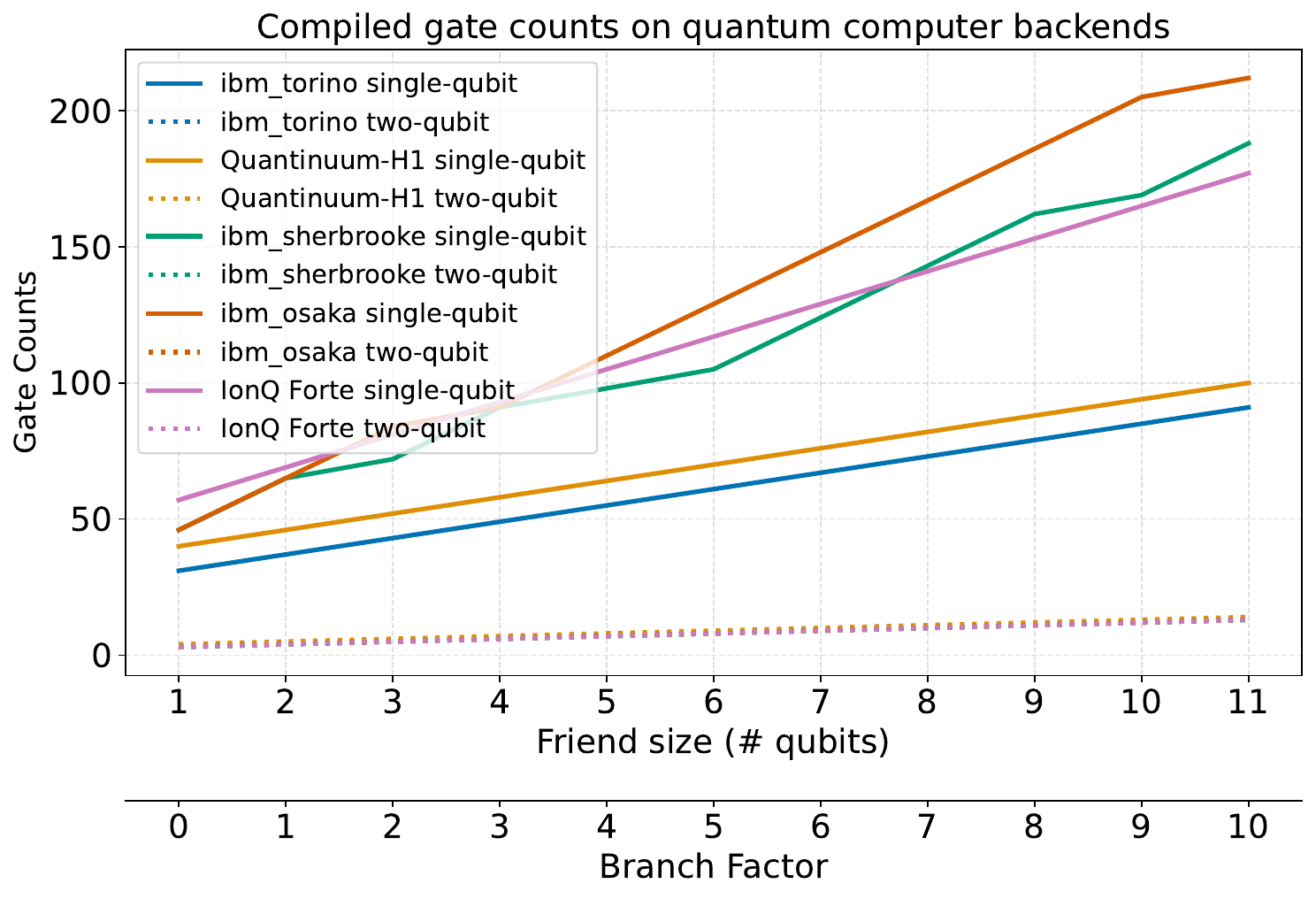}
        \label{fig:subfig2}
    \end{subfigure}
    \caption{(Left) A plot of the estimated fidelities of the friend state when
    prepared on different backends using the depolarizing model to estimate
    fidelity. The IBM Torino device supports valid LF violations up to branch
    factor 7, while the H1 device could theoretically support much larger
    violations. (Right) The single and two-qubit gate counts for compiled EWFS
    circuits on different backends.}
    \label{fig:validation}
\end{figure}

We use the methods in Section~\ref{sec:noisy-branches} to validate our
violations on real hardware. This section shows that we need confidence that our
friend state is prepared with a state fidelity of greater than 93.66\% to ensure
LF violations have been shown in the worst case. To estimate the friend's state
fidelity, we count the gates needed to prepare the state and use the per-gate
single and two-qubit error fidelities for the devices. IBM Torino has a single
and two-qubit gate errors that are sufficiently low to show violations up to
branch factor 9, while for the Quantinuum H1-1 device the errors are low enough
to allow for much higher branch factor violations, higher than branch factor 16
that we observed violations for. In Figure~\ref{fig:validation}, we plot the
estimated friend state fidelity vs. branch factor and the single and two-qubit
gate counts for various hardware backends.

%%%%%%%%%%%%%%%%%%%%%%%%%%%%%%%%%%%%%%%%%%%%%%%%%%%%%%
\section{More branch factor from less QPU: controlled-random unitary friends and Dicke states}
\label{sec:random-unitary-friends}
%%%%%%%%%%%%%%%%%%%%%%%%%%%%%%%%%%%%%%%%%%%%%%%%%%%%%%
In Section~\ref{sec:experiments}, we showed violations using only GHZ friends.
However, the quantum systems available for experiments have limited resources
and so this section considers tradeoffs in using those limited resources to
create high branch factors. This can inform future LF violation experiments. We
show theoretically and numerically that if you can use longer quantum circuits
for the friend states, you achieve higher branch factor with the same number of
qubits using Dicke states or controlled-random unitary states instead of GHZ
states.

We may want to use quantum states for the friend system, where the branch factor
grows quickly as we increase the number of qubits. In the case of the GHZ
friends, we saw in Equation~\eqref{eq:GHZ-bf} that the branch factor increases
linearly in the number of qubits. However, when we use a controlled-random
unitary, the branch factor increases exponentially in the number of qubits, c.f.
Equation~\eqref{eq:RU-bf}. Controlled-random unitary friends will obtain higher
branch factors for the same number of qubits. Implementing a controlled-random
unitary friend can be done by picking a Haar random unitary matrix and compiling
it to a controlled quantum circuit.

Inferring the branch of a controlled-random unitary friend is similar to the GHZ
case. We have two branches, the zero-state $\ket{0^n}$ and a Haar-random state
$\ket{\psi}$. In the noiseless case, we can distinguish with increasingly high
probability the two branches with just one measurement: if you measure a
non-zero bitstring, the branch is $\ket{\psi}$. If you measure $\ket{0^n}$, the
branch could have been $\ket{\psi}$, but the probability of this goes to zero
exponentially fast as $n$ grows. This is more challenging in the presence of
noise, mainly because the branch $\ket{0^n}$ might evolve to a superposition of
non-zero bitstrings. However, the highest amplitude should still be on the
$\ket{0^n}$ state, so with high probability, we will still correctly infer the
correct branch. The relevant observable in this case is, therefore, given by
\begin{equation}
    A = \ketbra{0^n}{0^n} - \sum_{x\neq 0^n}\ketbra{x}{x}.
\end{equation}
We observe violations by running the EWFS circuits in an ideal simulator using
the controlled-random unitary friends, which is observable for inferring
branches. With small depolarizing noise, we still obtain violations for a few
qubits; see Figure~\ref{fig:random_unitary_dicke_state_dep_noise}. However, it
looks to be challenging to show these same violations on today's quantum
hardware. The main reason is that the Haar random unitaries are very complex
because they require many single and two-qubit gates from some fixed gate-set
required from the hardware. Still, more sophisticated processors will eventually
be capable of delivering this performance.~\footnote{Note that supremacy
experiments~\cite{arute2019quantum} are similar but not quite the same as what
is needed. Here, we need not just Haar-random unitaries but also controlled
Haar-random unitaries. This increases the challenge for quantum processors.}

\begin{figure}[!htpb]
    \centering
    \includegraphics[scale=0.5]{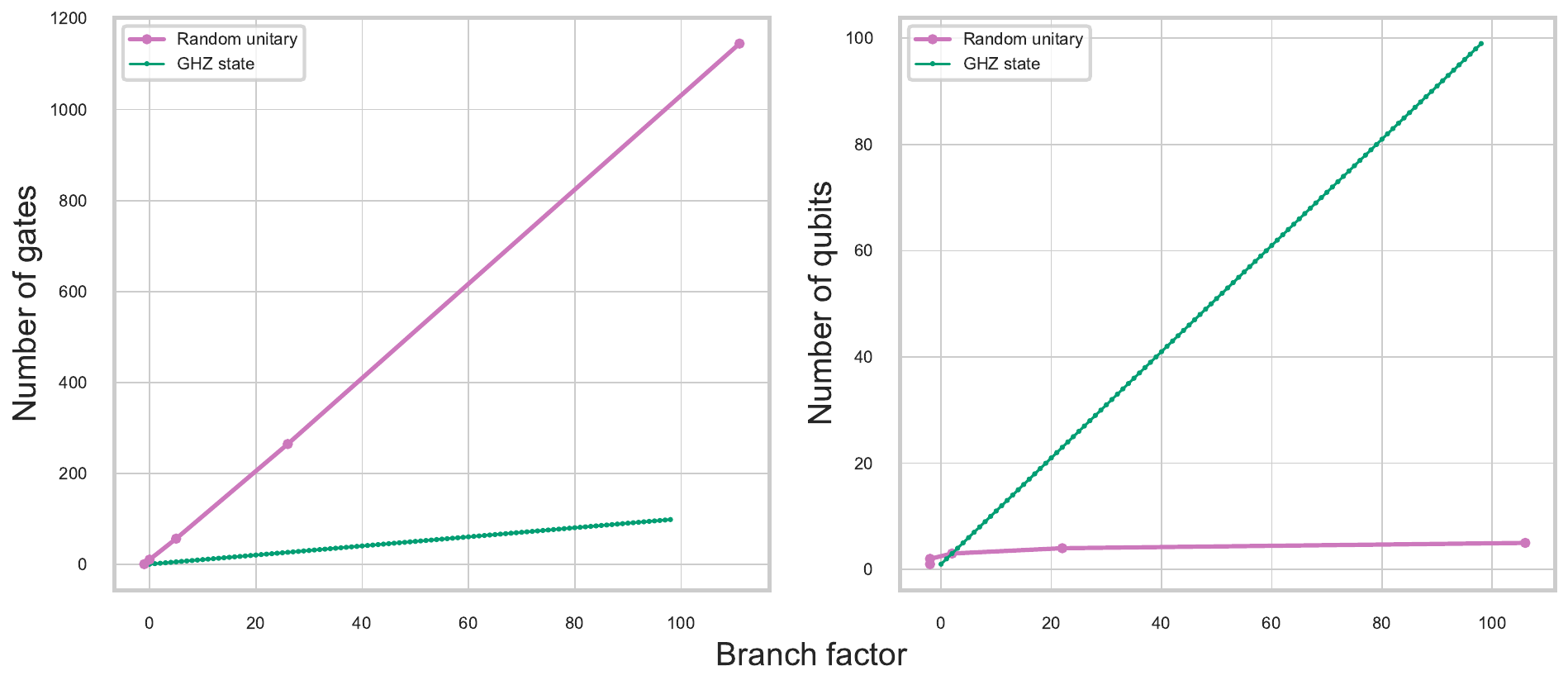}
    \caption{\textbf{Number of single and two-qubit gates and number of qubits
    required to prepare random unitary and GHZ states vs. branch factor.} For
    the basis gate set, we use the single qubit $X$, $Y$, and $Z$ rotations and
    CNOT gate, and we use Qiskit to transpile the given circuits. We see a
    tradeoff available for increasing branch factors depending on whether one is
    minimizing qubit number or gate count. Focusing on lower gate counts makes
    GHZ friends preferable while focusing on smaller numbers of qubits makes
    random unitary friends more preferable.}
    \label{fig:branchfactor_vs_gates}
\end{figure}

\begin{figure}[!htpb]
    \centering
    \includegraphics[scale=0.45]{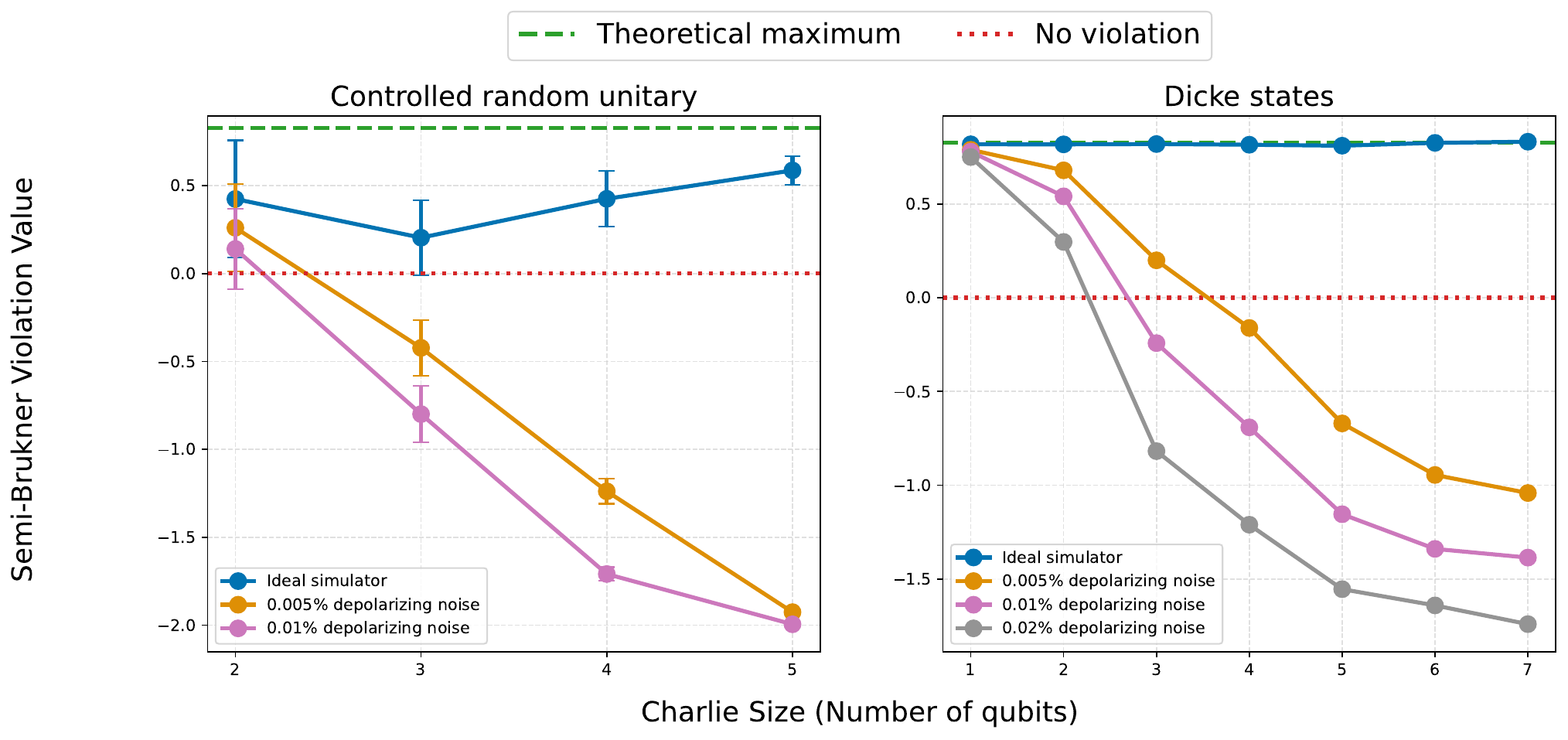}
    \caption{\textbf{Inequality violation values for the semi-Brukner inequality
    over an increasing range of depolarizing noise.} On the left, we consider a
    controlled-random unitary acting as the friends. On the right, we consider a
    controlled Dicke state as a friend. For the controlled-random unitary
    friends, we sample a Haar random unitary 100 times and take the average and
    error bars resulting from those runs. In the random unitary case, one sees
    that we don't obtain the optimal violation even in the ideal simulations.
    The reason is that inferring branches is probabilistic and fails with higher
    probability in the low friend sizes.}
    \label{fig:random_unitary_dicke_state_dep_noise}
\end{figure}

One way to obtain violations using such (complicated) circuits on noisy hardware
is to use states easily distinguishable from the all-zero state, even in the
presence of noise, while having a high swap complexity with the all-zero state.
For example, states with this property are Dicke
states~\cite{bartschi2019deterministic}. These are quantum states on $n$ qubits
with amplitudes only on bitstrings with a Hamming weight of $k$ for a choice of
$0\leq k \leq n$ denoted by $D(n,k)$. In~\cite{bartschi2019deterministic}, they
show quantum circuits with circuit complexity $O(kn)$. Assuming this is also a
lower bound, the equal superposition of such a state and the all-zero state
would have branch factor $\Omega(kn)$. For example, take $k=n/2$, so $D(n,n/2)$
is an equal superposition over bitstrings with Hamming weight $n/2$ and has
circuit complexity $O(n^2)$. In the presence of bitflip noise, it is unlikely
that too many bits get flipped to get close to the all-zero state. In this way,
even in the presence of noise, we are unlikely to measure the all-zero state.
Also, the noisy all-zero state will have amplitudes on non-zero bitstrings, but
these will probably have low Hamming weight. For example, the observable 
\begin{equation}
    A = \sum_{H(x)<n/3}\ketbra{x}{x} - \sum_{H(x)\geq n/3}\ketbra{x}{x}
\end{equation}
should be able to infer the right branch with high probability in the presence
of bitflip noise. Still, these states are too complex (with many single-qubit
gates and CNOTs) to run on noisy hardware and obtain violations. In
Figure~\ref{fig:random_unitary_dicke_state_dep_noise}, we show some violations
of the semi-Brukner inequality for small values of depolarizing noise. In
Figure~\ref{fig:branchfactor_vs_gates}, we plot the number of gates required to
implement a Haar random unitary and GHZ states on $n$ qubits versus the branch
factor. For the random unitary, we use the lower bound on the branch
factor~\ref{eq:RU-bf}, whereas for the GHZ states, we use the exact
value~\ref{eq:GHZ-bf}. We don't have a plot for Dicke states as that would
require an explicit lower bound, including constant factors on the circuit
complexity. However, we conjecture that it will lie between the GHZ state and
the random unitary resource requirements for both gates and qubits.

In Section~\ref{sec:validate-branch-factors}, we consider several types of
random quantum states that may be of interest because they have protocols (of
varying difficulty) to validate their preparation. Future work should consider
whether these states can also efficiently generate large branch factors with
limited qubit counts and gate numbers.

%%%%%%%%%%%%%%%%%%%%%%%%%%%%%%%%%%%%%%%%%%%%%%%%%%%%%%
\section{Future Directions}
\label{sec:future-directions}
%%%%%%%%%%%%%%%%%%%%%%%%%%%%%%%%%%%%%%%%%%%%%%%%%%%%%%
This work introduces a program for testing Local Friendliness using observers of
increasingly significant branch factors. We explain how quantum computers of
increasing power can run these Local Friendliness tests using Extended Wigner's
Friend scenarios. We then show violations of Local Friendliness at small scales
using currently available superconducting processors.

It is important to note that the LF violations shown in this work are not
\emph{loophole-free}. For example, we have not used the speed-of-light
limitations on communication to separate Alice and Charlie from Bob and Debbie.
In principle, loophole-free Local Friendliness violations could be performed
with networked quantum computers over a quantum
internet~\cite{wehner2018quantum}. Many proposals exist for building these
quantum links today, with companies and research groups investing in building
them at scale. We anticipate that, as quantum computers improve, quantum
networks connecting them will also improve. Further, EWFS only needs two nodes
to be connected by a quantum link. Thus, Local Friendliness violations are an
appealing fundamental science application for the next generations of quantum
computers and early quantum networks.

While this work ran EWFS on superconducting and ion-trap quantum processors
available over the cloud, other types of quantum processors, including neutral
atoms, silicon, and others, are being developed. As these platforms evolve, some
specific properties may be helpful for LF violations. For example, many neutral
atom platforms can natively execute global CZ
gates~\cite{wintersperger2023neutral, henriet2020quantum}. Combined with the
increasing number of qubits available in neutral atom platforms, neutrals could
be well-suited to producing the GHZ-type states we seek in larger and larger
friends.

Furthermore, we could consider designing quantum processing systems optimized
for producing LF violations at progressively larger branch factors. In a sense,
general-purpose quantum computers are overkill for running EWFS. We do not need
flexible re-programmability. Instead, we need to produce, for example, a
specific controlled-random unitary (or GHZ state) that we can reverse unitarily.
Perhaps a quantum ASIC or purpose-built quantum system can scale to a
significant branch factor with fewer engineering difficulties than needed for a
large fault-tolerant quantum computer. It may even be that EWFS, with more than
two friends and more choices of measurement settings for Alice and Bob, might be
advantageous in matching theory with easily buildable experiments. Designing
experimental systems can optimize performance both by increasing the branch
factor for fixed resources and also by preparing states whose branch factor can
be easily verified, as discussed in Section~\ref{sec:validate-branch-factors}.
This is a fresh direction for scaling up controlled quantum systems for
fundamental research.

Finally, there is an important outstanding theoretical question. What branch
factors should we aim for? Ideally, we would have milestones of specific branch
factor sizes corresponding to ruling in or out meaningful classes of physical
systems as observers. For example, what is the branch factor of a single photon
detector, the human eye, or the human brain? This would add meaningful threshold
milestones to experiments in the program of Figure~\ref{fig:gradient}. 

One approach is to work backward from physical systems that arguably most
resemble human observers. This is the approach taken
in~\cite{wiseman2023thoughtful}, where the authors consider a reversible
simulation of a human-level artificial intelligence running on a QPU as the
friend (QUALL-E). The authors estimate the size of a QPU needed to perform
experiments using a human-level artificial intelligence simulated on a QPU as
the friend observer at approximately $10^{19}$ logical qubits and logical depth
of $10^{14}$ operations. While we don't know the details of what branch factor
such a simulation might produce, we can estimate that a QPU capable of running
such a simulation could also produce GHZ states with a branch factor of order
$10^{19}$ as GHZ branch factors scale with qubit number. 

Alternatively, we can take inspiration from other experiments where the
quantum-classical cut is placed. In the loophole-free Bell
violations~\cite{giustina2015significant}, the experimenters needed to determine
at what point it was sufficient to declare the information classical in the
measurement chain. For reference, in~\cite{giustina2015significant} Figure 2
shows the spacetime diagram ending at the end of the detection period. After
this point, sufficient separation to ensure no loopholes was no longer needed.
This indicates that there is some accepted physical system that becomes
classical before it reaches a human. Calculating the corresponding branch factor
for this system would give a meaningful target. Future work can look to
calculate these branch factors for other meaningful systems to produce a road
map of increasingly meaningful Local Friendliness violations.

\subsection*{Code availability}

Software that implements the EWFS circuit and the code used to generate the data
and plots in this work are available on GitHub at~\cite{unitary2024ewfs}.

\subsection*{Acknowledgments}

We thank Eric Cavalcanti and Nora Tischler for discussing encoding the EWFS into
a quantum circuit, providing the data in the middle column of
Table~\ref{tab:ewfs-violations}. We additionally thank them, as well as Eleanor
Rieffel, Veronika Baumann, Andrea Mari, and anonymous reviewers for their
comments on the drafts of this manuscript. We thank Howard Wiseman, \v{C}aslav
Brukner, Felix Binder, and all the Wigner's Friends workshop attendees for
suggestions and discussion. VR thanks Kevin Sung, Jake Lishman, and Luciano
Bello for discussions on Qiskit and running on IBM hardware. We also thank the
Nexus team for guiding our experiments on the Quantiuum H1 device. VR and FL
thank Matthew Beach for extensive debugging and assistance in running
experiments via the Braket platform. WZ thanks the O'Shaughnessy Fellowship for
support of this work. This work was supported in part by ARC Grant No.
DP250102162.
  
%-----------------------------------------------------% 
\bibliographystyle{quantum} 
\bibliography{references} 
%-----------------------------------------------------%

\appendix

%%%%%%%%%%%%%%%%%%%%%%%%%%%%%%%%%%%%%%%%%%%%%%%%%%%%%%%%%%%%%%%%%%%%%%%%%%%%%%%%%
\section{Derivation of Semi-Brukner inequality without Debbie}
\label{sec:semi-brukner-no-debbie-derivation}
%%%%%%%%%%%%%%%%%%%%%%%%%%%%%%%%%%%%%%%%%%%%%%%%%%%%%%%%%%%%%%%%%%%%%%%%%%%%%%%%%
\noindent We want to prove the inequality 
\begin{equation}
    S:=-\expval{A_1 B_2} + \expval{A_1 B_3} - 
    \expval{A_3 B_2} - \expval{A_3 B_3} \leq 2, 
\end{equation}
is obeyed when the assumptions of Absoluteness of Observed Events and Local
Agency are assumed. Assuming AOE and LA implies that the distribution of
outcomes $a,b$ for Alice and Bob given measurements choices $x$ and $y$
respectively implies (See Sec~\ref{sec:semi-brukner-no-debbie}):
\begin{equation}
    \phi(ab|xy) =
    \begin{cases}
        \sum_{c} \delta_{a,c}P(b|cy)P(c) & \text{if } x = 1 \\
        \sum_{c} P(ab|cxy)P(c) & \text{if } x \neq 1
    \end{cases}
\end{equation}
where $a,b,c \in \{-1,1\}$, $\expval{A_x B_y} := \sum_{a,b} ab \phi(ab|xy)$.
Expanding each term in the expression for $S$ using the correlations defined
above
\begin{equation}
    S = \sum_{c} P(c) \left[ -c \expval{B_2}_c + c \expval{B_3}_c - 
    \expval{A_3 B_2}_c - \expval{A_3 B_3}_c \right]
\end{equation}
where we define $\expval{B_y}_c = \sum_{b}bP(b|cy)$ and $\expval{A_x B_y}_c =
\sum_c ab P(ab|cxy)$. Let 
\begin{equation}
    X(c) = -c \expval{B_2}_c + c \expval{B_3}_c - \expval{A_3
    B_2}_c - \expval{A_3 B_3}_c.
\end{equation}
If we can show that $X(c) \leq 2$ for all $c \in \{-1,1\}$, then $S \leq \sum_c
P(c) \cdot 2 = 2$, since $\sum_c P(c) = 1$. Since $P(b|cy) = \sum_a P(ab|cxy)$
we have that $X(c)$ can be written as
\begin{equation}
    X(c) = c \sum_{a,b} b P(ab|c\,3\,3) - c \sum_{a,b} b
    P(ab|c\,3\,2) - \sum_{a,b} ab P(ab|c\,3\,2) - \sum_{a,b} ab
    P(ab|c\,3\,3).
\end{equation}
Here we use that the marginal $P(b|cy)$ is independent of $x$ using the Local
Agency assumptions (Sec~\ref{sec:semi-brukner-no-debbie}). Rearranging the terms
based on the $(xy)$ settings for $P(ab|cxy)$
\begin{equation}
    X(c) = \sum_{a,b} b(c-a) P(ab|c\,3\,3) - \sum_{a,b} b(c+a) P(ab|c\,3\,2).
\end{equation}
Now analyze the terms $c-a$ and $c+a$. Since $a, c \in \{-1,1\}$, $X(c)$ will
only be maximized when both $(c-a)$ and $(c+a)$ are non-zero
\begin{itemize}
    \item $c-a$ is non-zero only if $a = -c$. In this case, $c-a = c - (-c) = 2c$.
    \item $c+a$ is non-zero only if $a = c$. In this case, $c+a = c + c = 2c$.
\end{itemize}
Thus $X(c)$ can be maximized by setting $(c-a)=(c+a)=2c$. The expression for
$X(c)$ then simplifies to
\begin{equation}
    X(c) = 2c \left[ \sum_b b P(a=-c, b | c\,3\,3) - \sum_b b P(a=c, b | c\,3\,2) \right].
\end{equation}
Let $S_1(c) = \sum_b b P(a=-c, b | c\,3\,3)$ and $S_2(c) = \sum_b b P(a=c, b |
c\,3\,2)$. Then, 
\begin{equation}
    X(c) = 2c (S_1(c) - S_2(c)).
\end{equation}
Using the triangle inequality, we have $$|S_1(c)| \leq \sum_{b}P(a=-c, b |
c\,3\,3) = P(a=-c | c\,3\,3).$$ Similarly, $|S_2(c)| \leq P(a=c | c\,3\,2)$.
Since $|c|=1$, again by the triangle inequality
\begin{equation}
    X(c) \leq 2 |S_1(c) - S_2(c)| \leq 2 (|S_1(c)| + |S_2(c)|).
\end{equation}
Using the bounds for $|S_1(c)|$ and $|S_2(c)|$:
\begin{equation}
    X(c) \leq 2 \left[P(a=-c | c\,3\,3) + P(a=c | c\,3\,2) \right].
\end{equation}
Local Agency implies that Alice's marginal probability $P(a|cx)$ (obtained by
summing $P(ab|cxy)$ over $b$) is independent of Bob's setting $y$. So, $P(a=-c |
c\,3\,3) = P(a=-c | c\,3)$ and $P(a=c | c\,3\,2) = P(a=c | c\,3)$. Thus, 
\begin{equation}
    X(c) \leq 2 \left[ P(a=-c | c\,3) + P(a=c |
    c\,3) \right].
\end{equation}
Since $a$ can only take values $c$ or $-c$ (because $a, c \in \{-1,1\}$), the
sum of these probabilities is $P(a=-c | c\,3) + P(a=c | c\,3) = 1$. Therefore,
$X(c) \leq 2 \cdot 1 = 2$.

Since $X(c) \leq 2$ for all possible values of $c$, and $P(c)$ is a probability
distribution, we have
\begin{equation}
    S = \sum_{c} P(c) X(c) \leq \sum_{c} P(c) \cdot 2 = 2.
\end{equation}

%%%%%%%%%%%%%%%%%%%%%%%%%%%%%%%%%%%%%%%%%%%%%%%%%%%%%%%%
\section{Running EWFS on quantum hardware}
\label{app:running-ewfs-on-quantum-hardware}
%%%%%%%%%%%%%%%%%%%%%%%%%%%%%%%%%%%%%%%%%%%%%%%%%%%%%%%%
To run the EWFS circuit on a hardware backend, the circuit must be transpiled to
target the hardware architecture. Transpiling the circuit in Qiskit is achieved
via the \texttt{transpile} function. This function takes an optionally specified
\texttt{optimization\_level} argument on how much optimization to perform on the
circuit. This value equals 1 by default (with a maximum value of 3 and a minimum
of 0), in which the value of 1 applies a light optimization across the overall
circuit. As the \texttt{optimization\_level} argument increases, more aggressive
strategies are applied to the circuit via transpilation to reduce gates in the
circuit and optimally route qubit mappings. In our case, we wish to avoid
gate-collapsing optimizations and instead focus on ideal routing. To achieve
this, we set \texttt{optimization\_level=0} to avoid any gate reductions. All
software used to conduct experiments, generate plot figures, and process data
from both hardware and simulator devices is available on
GitHub~\cite{unitary2024ewfs}.

%%%%%%%%%%%%%%%%%%%%%%%%%%%%%%%%%%%%%%%%%%%%%%%%%%%%%%%%
\subsection{Running EWFS on IBM quantum hardware}
\label{app:running-ewfs-on-ibm-quantum-hardware}
\begin{figure}[!htpb]
    \centering
    \includegraphics[scale=0.12]{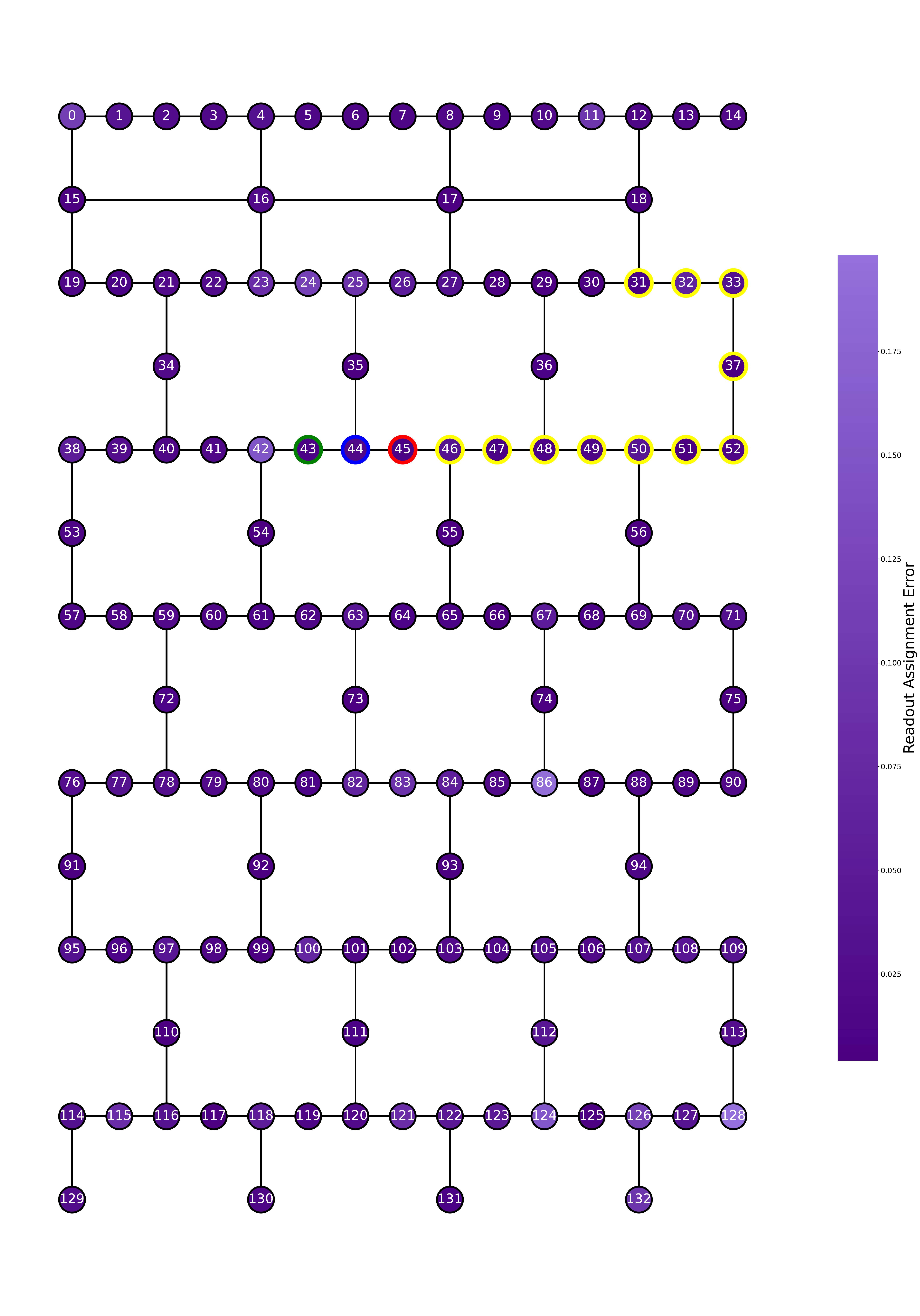}
    \caption{The calibrations coupling map for the CNOT readout error of CNOT
    gates for the \texttt{ibm\_torino} device. Lighter colors indicate a higher
    CNOT error, while darker shades indicate a lower CNOT error. When targeting
    this device, we aim to maintain a chain-like orientation for our initial
    layout and also pick a layout that optimizes for the lowest readout error.
    Specifically, we assign nodes labeled 43 to Debbie (highlighted in green),
    45 to Alice (highlighted in red), and 44 to Bob (highlighted in blue). The
    remaining nodes [46, 47, 48, 49, 50, 51, 52, 37, 33, 32, 31] (highlighted in
    yellow)represent the size of Charlie's system for progressively higher qubit
    systems. The nodes with the outer layer highlight indicate the qubit layout
    used for our experiments. This image was produced with device readout
    assignment error data obtained via the IBM Quantum Platform on 8-3-2024.}
    \label{fig:ibmq-torino-coupling-map}
\end{figure}

%%%%%%%%%%%%%%%%%%%%%%%%%%%%%%%%%%%%%%%%%%%%%%%%%%%%%%%%
We hand-optimize a routing layout position for the virtual qubits and physical
device qubits. We perform this by supplying an argument of
\texttt{initial\_layout} to the \texttt{transpile} method. We construct the
layout pattern based on the hardware coupling map and opt for attaining a
chain-like connectivity of qubits for Alice, Bob, Charlie, and Debbie. An
example of such a layout construction on the \texttt{ibm\_torino} device is
shown in Figure~\ref{fig:ibmq-torino-coupling-map}, but the same layout
orientation strategy was followed for the other IBM hardware devices as well. As
an additional step in our transpilation pipeline, we optimize single-qubit gate
decompositions using the \texttt{Optimize1qGatesDecomposition} class via the
\texttt{PassManager} in Qiskit.

% This figure was generated via networkx using the readout error data captured
% from the IBM platform.

%-----------------------------------------------------% 
\subsection{Running EWFS on Quantinuum H1-1 quantum hardware}
\label{app:running-ewfs-on-quantinuum-quantum-hardware}
%-----------------------------------------------------% 
Unlike the IBM hardware we target, the routing optimizations do not need to be
considered, as the H1-1 device coupling map corresponds to a fully connected
20-node complete graph. This means there is no need to hand-optimize an optimal
routing layout since every node is connected to every other node. We encoded our
EWFS circuits in Qiskit (as we did for IBM hardware) and transpiled these
circuits using TKET~\cite{tket} to target the Quantinuum H1 architecture. To run
our experiments, we used the Nexus~\cite{quantinuum_nexus} platform.

\end{document}